\documentclass[11pt]{article}
\pdfoutput=1

\usepackage{amsmath,amssymb,amsfonts,graphicx,amsfonts}
\usepackage{textcomp,mathcomp}
\usepackage{xcolor}
\usepackage[hidelinks]{hyperref}
\usepackage{ascmac}
\usepackage{physics}
\usepackage{cite}
\usepackage{comment}

\allowdisplaybreaks[4]

\date{}
\begin{document}
\begin{flushright}
YITP-23-121
\\
%\today\\
\end{flushright}
%\preprint{YITP-23-121}
\vspace{0.1cm}

\begin{center}
 {\Large A new perspective on thermal transition in QCD}\\

\end{center}
\vspace{0.1cm}
\vspace{0.1cm}
\begin{center}

Masanori Hanada$^a$, 
Hiroki Ohata,$^b$ 
Hidehiko Shimada,$^{b,c}$ 
and Hiromasa Watanabe$^b$

\end{center}
\vspace{0.3cm}

\begin{center}

{\small

$^a$School of Mathematical Sciences, Queen Mary University of London\\
Mile End Road, London, E1 4NS, United Kingdom

$^b$Yukawa Institute for Theoretical Physics, Kyoto University\\
Kitashirakawa Oiwakecho, Sakyo-ku, Kyoto 606-8502, Japan

$^c$
National Institute of Technology, Akashi College\\
679-3 Nishioka, Uozumi-cho, Akashi, Hyogo, 674-8501 Japan

}
\end{center}

\vspace{0.5cm}

\begin{center}
  {\bf Abstract}
\end{center}

Motivated by the picture of partial deconfinement developed in recent years for large-$N$ gauge theories, we propose a new way of analyzing and understanding thermal phase transition in QCD. We find nontrivial support for our proposal by analyzing the WHOT-QCD collaboration's lattice configurations for SU(3) QCD in $3+1$ spacetime dimensions with up, down, and strange quarks. 

We find that the Polyakov line (the holonomy matrix around a thermal time circle) is governed by the Haar-random distribution at low temperatures. The deviation from the Haar-random distribution at higher temperatures can be measured via the character expansion, or equivalently, via the expectation values of the Polyakov loop defined by the various nontrivial representations of SU(3). 

We find that the Polyakov loop corresponding to the fundamental representation and loops in the higher representation condense at different temperatures. This suggests that there are three phases, one intermediate phase existing in between the completely-confined and the completely-deconfined phases. Our identification of the intermediate phase is supported also by the condensation of instantons: by studying the instanton numbers of the WHOT-QCD configurations, we find that the instanton condensation occurs for temperature regimes corresponding to what we identify as the completely-confined and intermediate phases, whereas the instantons do not condense in the completely-deconfined phase. 

Our characterization of confinement based on the Haar-randomness explains why the Polyakov loop is a good observable to distinguish the confinement and the deconfinement phases in QCD despite the absence of the $\mathbb{Z}_3$ center symmetry. 

\newpage

%%%%%%%%%%%%
%%%%%%%%%%%%
\section{Introduction}\label{sec:introduction}
\hspace{0.51cm}
%%%%%%%%%%%%
%%%%%%%%%%%%
Confinement/deconfinement transition in gauge theory~\cite{Polyakov:1978vu, Susskind:1979up} is an important phenomenon that has applications ranging from the quark-gluon plasma phase in QCD under extreme conditions to the description of black holes via gauge/gravity duality. 
However, the definition of confinement and deconfinement has been somewhat unclear in real-world QCD with three colors. The biggest issue was the lack of the strict notion of symmetry characterizing confinement and deconfinement. 
Specifically, the $\mathbb{Z}_3$ center symmetry that provides us with a good characterization for pure Yang-Mills theory (i.e., unbroken and broken center symmetry correspond to confinement and deconfinement, respectively) does not exist for QCD due to the presence of quarks in the fundamental representation.
The Polyakov loop\footnote{We will call the holonomy matrix (in the fundamental representation) along a thermal circle as the Polyakov line. Its trace is the standard Polyakov loop. We will also consider the trace of the holonomy in representations other than the fundamental representations. We will also call them the Polyakov loops in these representations.}
no longer plays the role of the order parameter associated with the center symmetry. Nonetheless, the Polyakov loop has been empirically used as an ``order parameter'' of
the deconfinement transitions.

Investigations of large-$N$ gauge theories have been successful in elucidating the nontrivial nature of QCD such as the occurrence of deconfinement transition. 
It has been known that the phase transition can be detected by the Polyakov line (i.e., the holonomy matrix whose trace gives the Polyakov loop) in an independent way to center symmetry.~\footnote{This was analytically shown for weakly-coupled theories on a three-sphere~\cite{Sundborg:1999ue, Aharony:2003sx, Schnitzer:2004qt}. Even for theories with center symmetry, a phase transition sits between two center-broken phases.} The reason that the Polyakov line captures the phase transition turned out to be its connection to gauge symmetry~\cite{Hanada:2020uvt}. 

In this letter and a companion paper~\cite{Hanada:2023rlk}, we explore whether and how this idea in large-$N$ theory can be applied to the real-world QCD, i.e., SU($N=3$) QCD with dynamical quarks. 
Note that, although we are considering finite $N$, the phase transition is not prohibited because the thermodynamic limit is realized at infinite volume. 

We take a bottom-up approach in this letter, by looking into lattice data generated by WHOT-QCD collaboration~\footnote{
WHOT-QCD collaboration \cite{Umeda:2012er} studies the finite temperature QCD using the Wilson fermion. Although the mass parameters of fermions of WHOT-QCD are not small enough to reproduce the correct meson spectrum, they are small enough so that there is no first-order thermal phase transition, which is expected to be the behavior of QCD at physical quark mass\cite{Aoki:2006br, Aoki:2006we}.
} in Sec.~\ref{sec:lattice_data} and then giving the interpretation based on the connection to the theoretical understanding of large-$N$ theory in Sec.~\ref{sec:comparison_with_large_N}. The presentation in the companion paper~\cite{Hanada:2023rlk} follows a top-down approach, i.e., we start with the large-$N$ theories, make conjectures on SU(3) QCD, and then confirm these conjectures based on lattice QCD data. In fact, we took such a top-down approach in our investigation.

%%%%%%%%%%%%
%%%%%%%%%%%%
\section{Looking into lattice data}\label{sec:lattice_data}
\hspace{0.51cm}
%%%%%%%%%%%%
%%%%%%%%%%%%
We consider thermal circles defined at each spatial point $\vec{x}$ and the Polyakov lines $P_{\vec{x}} \in {\rm SU}(3)$.
A technical but crucial idea which facilitates our analysis is to consider an ensemble (probability distribution)
of the Polyakov line, counting $P_{\vec{x}}$ for each $\vec x$ as a sample.~\footnote{
In this letter, we give estimations of statistical errors ignoring the spatial correlation of the Polyakov loops at different spatial points. (Investigation of the WHOT-QCD data shows that this gives a good estimate of the errors.)
}

In Table~\ref{table:WHOT}, we show a short profile of the WHOT-QCD lattice configurations at seven different values of temperature which we use, together with our identification of the phases associated with the temperatures which will be explained below. 
The spatial volume is $32^3$, and hence, we obtain $32^3=32768$ $P_{\vec{x}}$'s from each lattice QCD configuration.\footnote{
Different temperatures are obtained by using the same lattice spacing and different lattice sizes. This fixed-UV-scale approach justifies the use of the bare Polyakov loops, as we will discuss more in the discussion section.}

\begin{table}[hbtp]
  \centering
  \begin{tabular}{|c|c|c|}
    \hline
    Lattice size  & Temperature & Phase\\
    \hline 
    $4\times 32^3$   &  697 MeV & \textcolor{red}{CD}\\
    $6\times 32^3$   &  464 MeV & \textcolor{red}{CD}\\
    $8\times 32^3$   &  348 MeV & \textcolor{orange}{PD} or \textcolor{red}{CD}\\
    $10\times 32^3$  &  279 MeV & \textcolor{orange}{PD} \\
    $12\times 32^3$  &  232 MeV & \textcolor{orange}{PD}\\
    $14\times 32^3$  &  199 MeV & \textcolor{orange}{PD}\\
    $16\times 32^3$  &  174 MeV & \textcolor{blue}{CC} or \textcolor{orange}{PD}\\  
    \hline
  \end{tabular}
    \caption{WHOT-QCD configurations of
    $N_{\rm f}=2+1$ QCD~\cite{Umeda:2012er}. Lattice spacing $a\simeq 0.07$ fm, with up and down quarks heavier than physical mass (i.e., $\frac{m_\pi}{m_\rho}\simeq 0.63$ is larger than the real-world value $\frac{m_\pi}{m_\rho}\simeq 0.18$) and approximately physical strange-quark mass. The third column is our estimate of the phase discussed in the main text. 
    \textcolor{red}{CD}, \textcolor{orange}{PD}, and \textcolor{blue}{CC} denote 
    \textcolor{red}{Complete Deconfinement}, 
    \textcolor{orange}{Partial Deconfinement}, and \textcolor{blue}{Complete Confinement}. 
}\label{table:WHOT}
\end{table}

\begin{figure}[htbp]
\begin{center}
\scalebox{0.33}{
\includegraphics{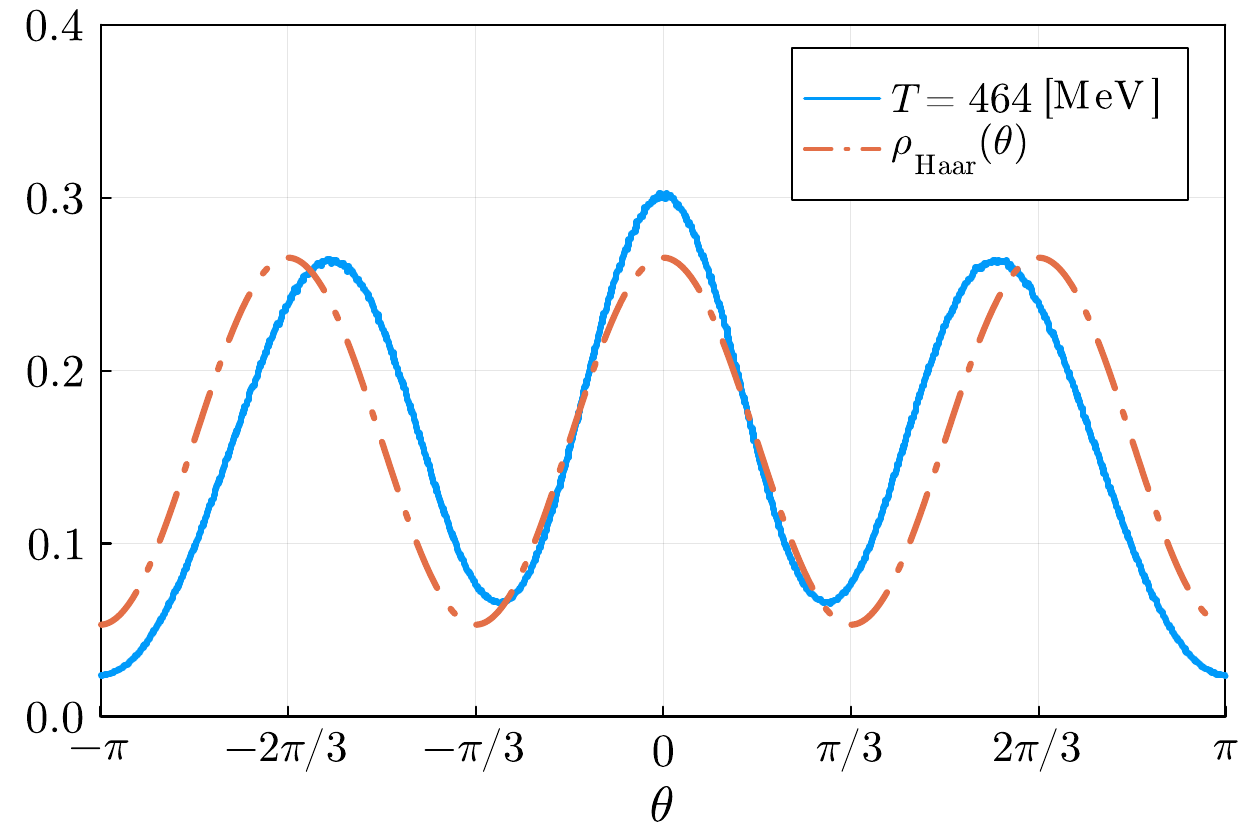}}
\scalebox{0.33}{
\includegraphics{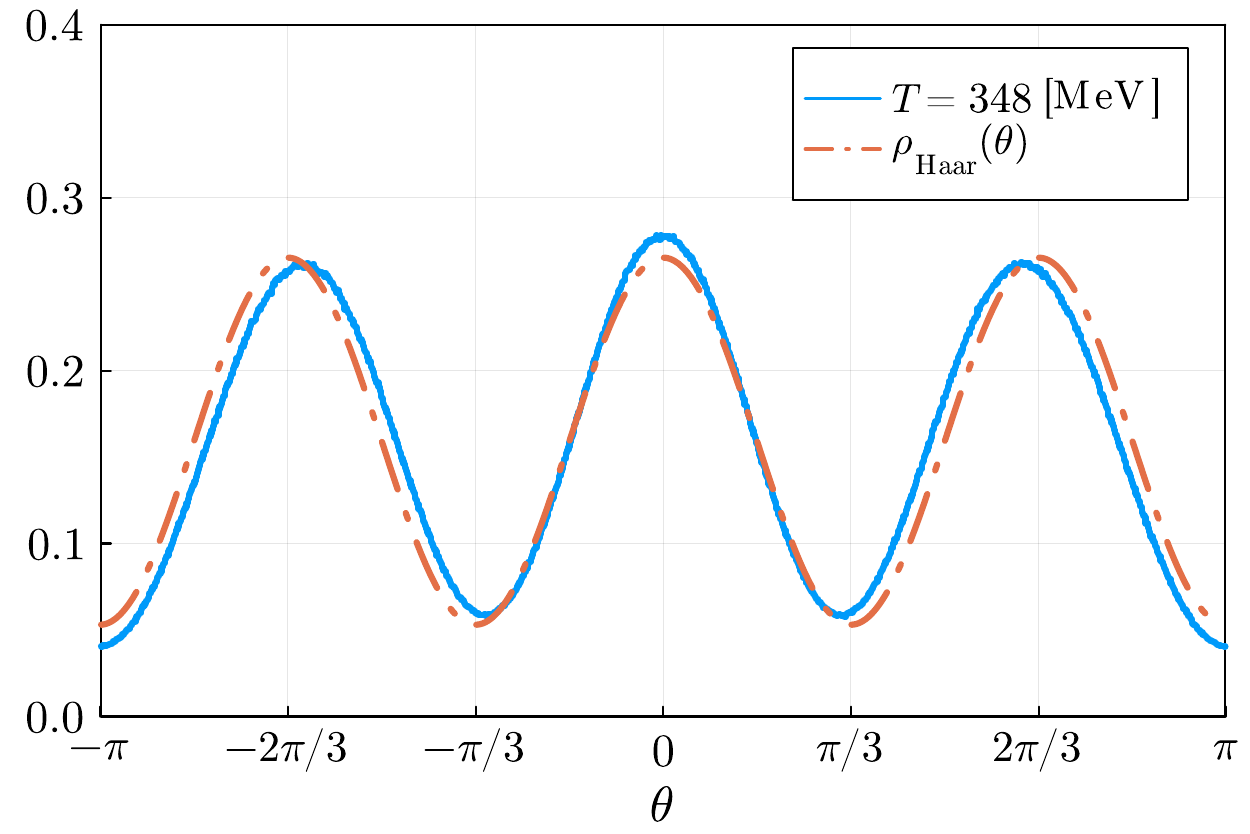}}
\scalebox{0.33}{
\includegraphics{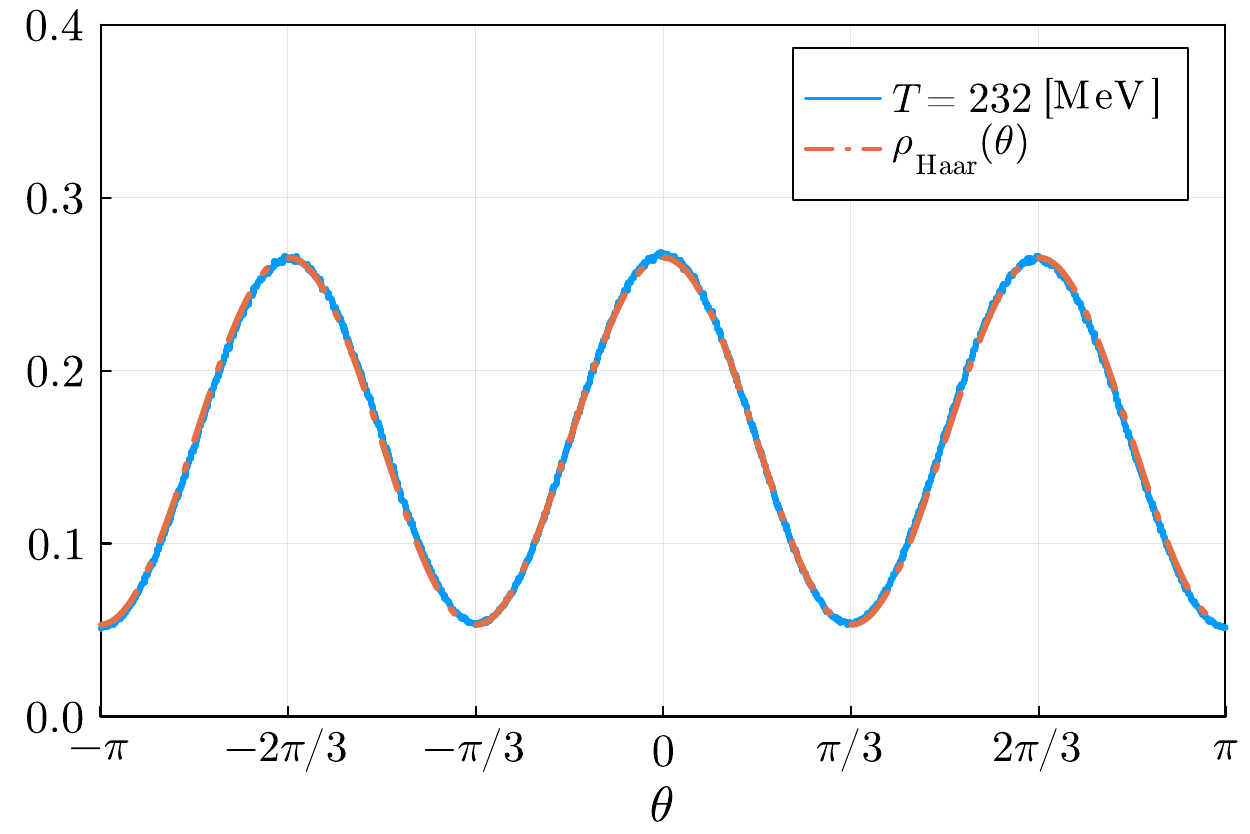}}
\scalebox{0.33}{
\includegraphics{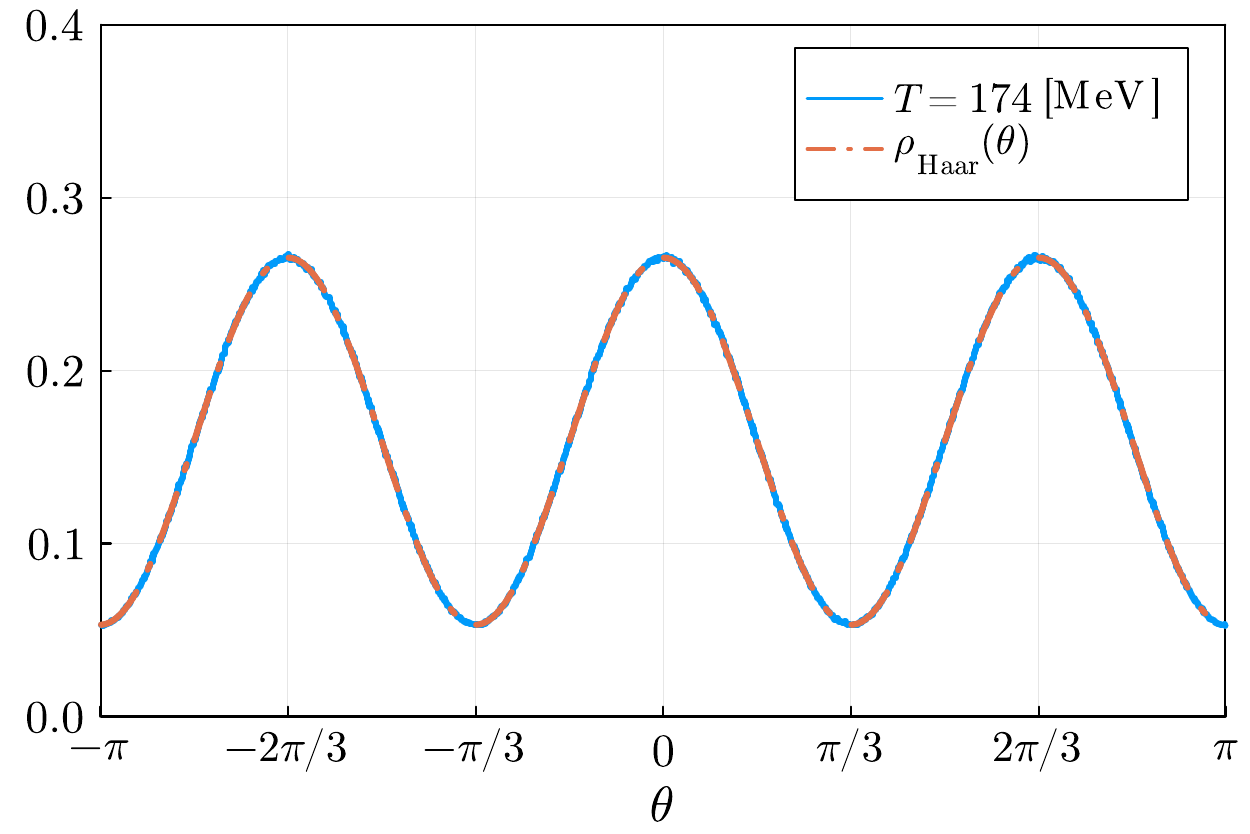}}
\end{center}
\caption{Distribution of Polyakov line phases, $N_t \times 32^3$ lattice, obtained from the WHOT-QCD configurations, is shown as the solid line.
The lattice size is $N_t\times 32^3$ ($N_t=6,8,12,16$, and correspondingly $T =$ 464, 348, 232, and 174 MeV.) and 599 configurations were used. 
These histograms are drawn with 992 bins. Although the agreement with the Haar-random distribution shown in the dash-dot line (see eq.~\eqref{dist:Haar-random}) seems to be good at $T=232$~MeV, more careful investigation reveals a small deviation and hence the onset of partial deconfinement; see Figs.~\ref{fig:T-vs-character} and \ref{fig:T-vs-fund_character} and main text.
At 174 MeV, agreement with the Haar-random distribution is much better. 
}
\label{fig:WHOT-phase}
\end{figure}

At each spatial point $\vec{x}$, $P_{\vec{x}}$ is a $3\times 3$ matrix with eigenvalues $e^{i\theta_1}, e^{i\theta_2}$ and $e^{i\theta_3}$, with $\theta_1+\theta_2+\theta_3\equiv 0$ mod $2\pi$. There are $3\times 32^3=98304$ eigenvalues per configuration. We can estimate the distribution $\rho(\theta)$ ($-\pi<\theta\le\pi$) by combining many configurations. The results are shown in Fig~\ref{fig:WHOT-phase}. We compare this distribution against that would arise from the Haar-random distribution on $\rm{SU}(3)$, 
\begin{align}
\rho_{\rm Haar}(\theta)=\frac{1}{2\pi}\left(1+\frac{2}{3}\cos(3\theta)\right).
\label{dist:Haar-random}
\end{align}
(For completeness, we give the derivation of this formula in appendix \ref{sec:Haar-random-distribution}.) 
The plots show that $\rho(\theta)$ deviates from the Haar-random distribution for higher temperatures $T=348~{\rm MeV}$, $ 464~{\rm MeV}$, but appears indistinguishable for lower temperatures, with our naked eyes, from $\rho_{\rm Haar}(\theta)$ at $T = 174~{\rm MeV}, 232~{\rm MeV}$. 
We will discuss a method to measure the deviation from the Haar-random distribution systematically shortly below, which shows that the deviation decreases rapidly toward $T = 174~{\rm MeV}$.

The agreement with the Haar-random distribution at low temperatures is a crucial feature that has been theoretically understood in the large-$N$ theories based on analysis focused on the redundancy of the states under gauge transformations.~\footnote{It might not come as a surprise that the Polyakov line obeys the Haar-randomness at strictly zero temperature; at $T=0$ the Polyakov line is a product of infinitely many link variables, and because of this the probability distribution of the Polyakov line may converge into the Haar-random distribution. What is quite remarkable is that the Polyakov line obeys the Haar-random distribution to good accuracy even at rather high temperatures right below the deconfinement transition.}
We provide a short summary of this point in Appendix \ref{sec:meaning-Polyakov-large-N}.  

We can measure the deviation from the Haar-randomness quantitatively by using character expansion.~\footnote{
We note that Polyakov, in the pioneering work~\cite{Polyakov:1978vu}, pointed out the usefulness of character expansion to characterize the  
deconfinement transition. Apparently, 
this prescient remark was not followed up seriously
before our work.
}
(We collect some properties of character expansion together with explicit formulae for characters of $\rm{SU}(3)$ in Appendix \ref{sec:character}.)
We write the probability distribution of the Polyakov line $P \in \rm{SU}(3)$ by $\rho(P)$. Let $\chi_r(P)$ be the character associated with an irreducible representation $r$. By the completeness of the characters, one can expand $\rho(P)$ in terms of $\chi_r(P)$ as $\rho(P)=\sum_r\rho_r \chi_r(P)$, where the expansion coefficients are $\rho_r = \int dP \rho(P) \chi_r^*(P)$ because of the orthonormality of the characters $\int dP \chi_r(P) \chi_{r'}^*(P)=\delta_{rr'}$. By construction, $\rho_r$ coincides with the expectation value of the Polyakov loop in the representation $r$.

For the exact Haar-random distribution, $\rho(P)$ is completely dominated by the trivial representation, i.e., $\rho_r$ vanishes for any nontrivial representation $r$. Hence, the Polyakov loops in nontrivial representations give good measures of the deviation of the Polyakov line phases from the Haar-random distribution. Note that $\rho_r$ contains all statistical information of the distribution of the Polyakov line, including the correlations between three eigenvalues of the Polyakov line such as the level repulsion.

\begin{figure}[htbp]
\begin{center}
\scalebox{0.33}{
\includegraphics{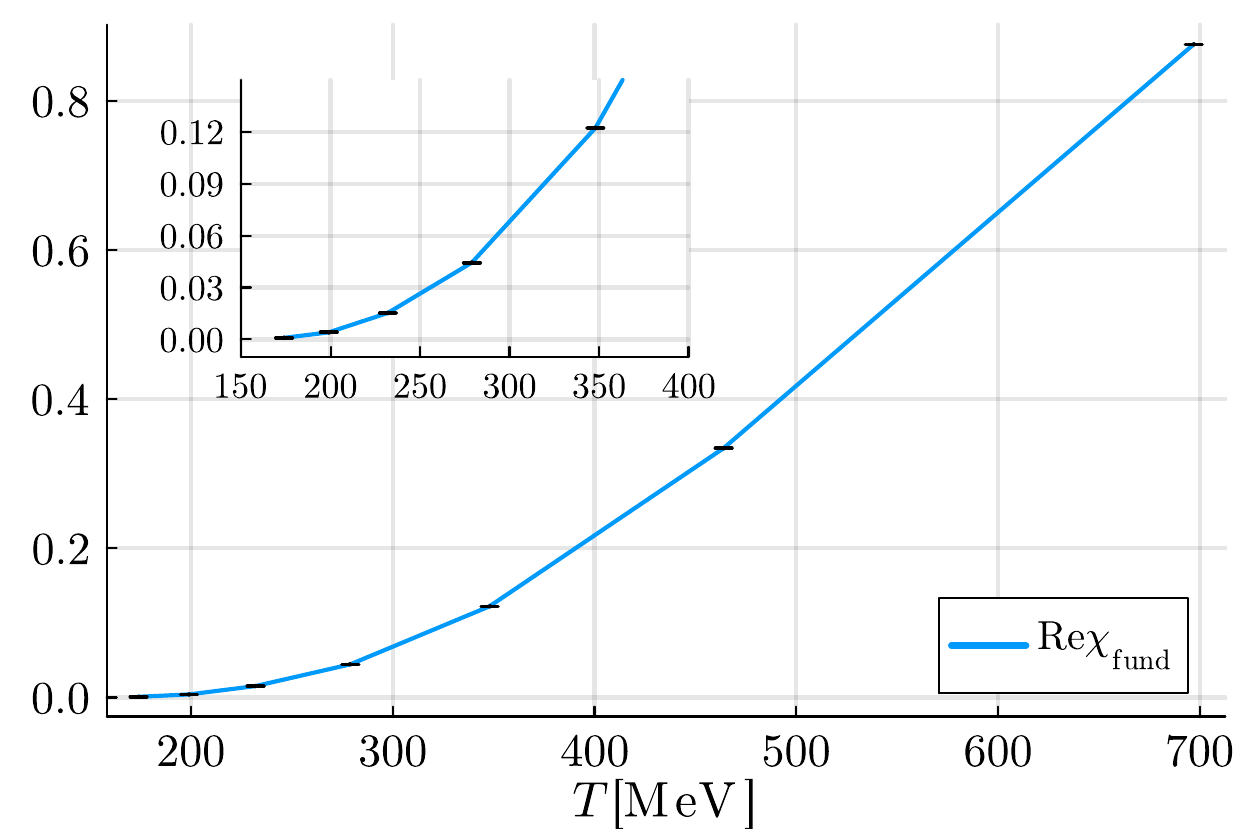}}
\scalebox{0.33}{
\includegraphics{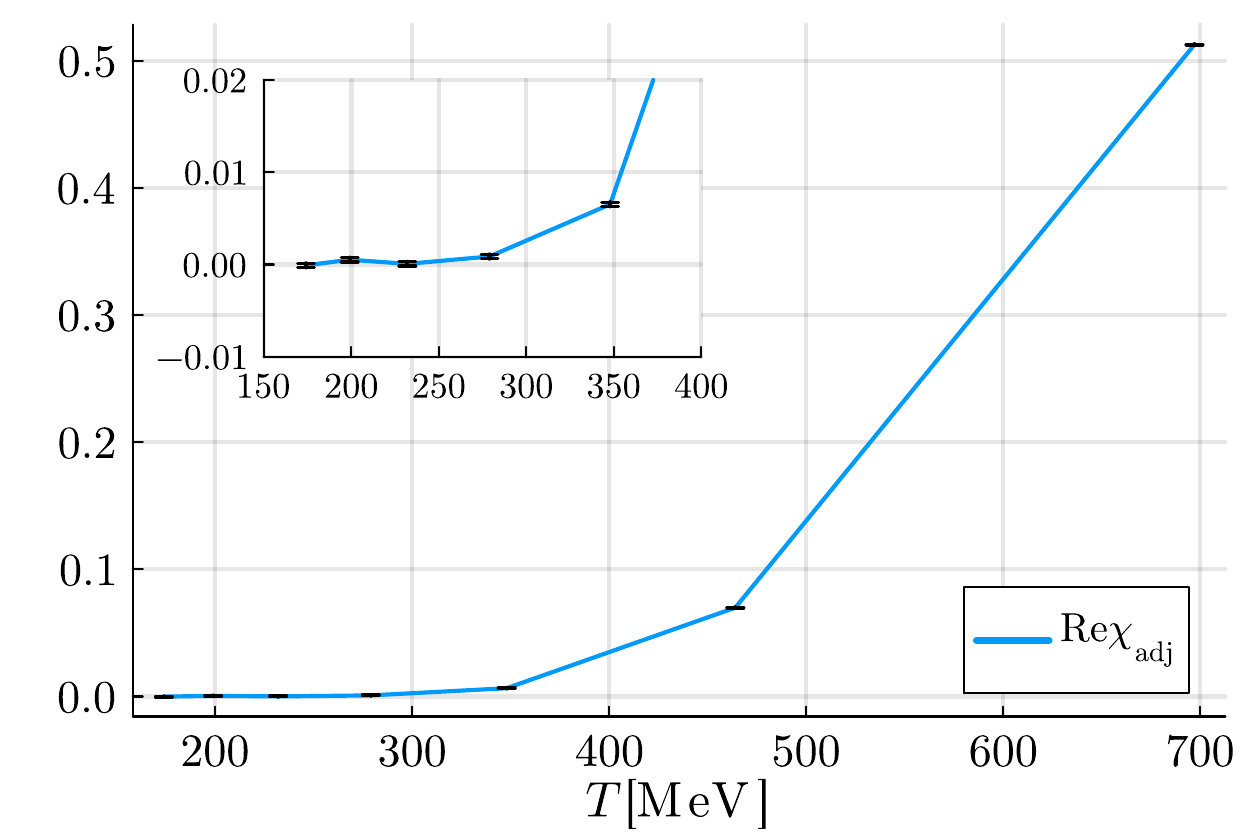}}
\scalebox{0.33}{
\includegraphics{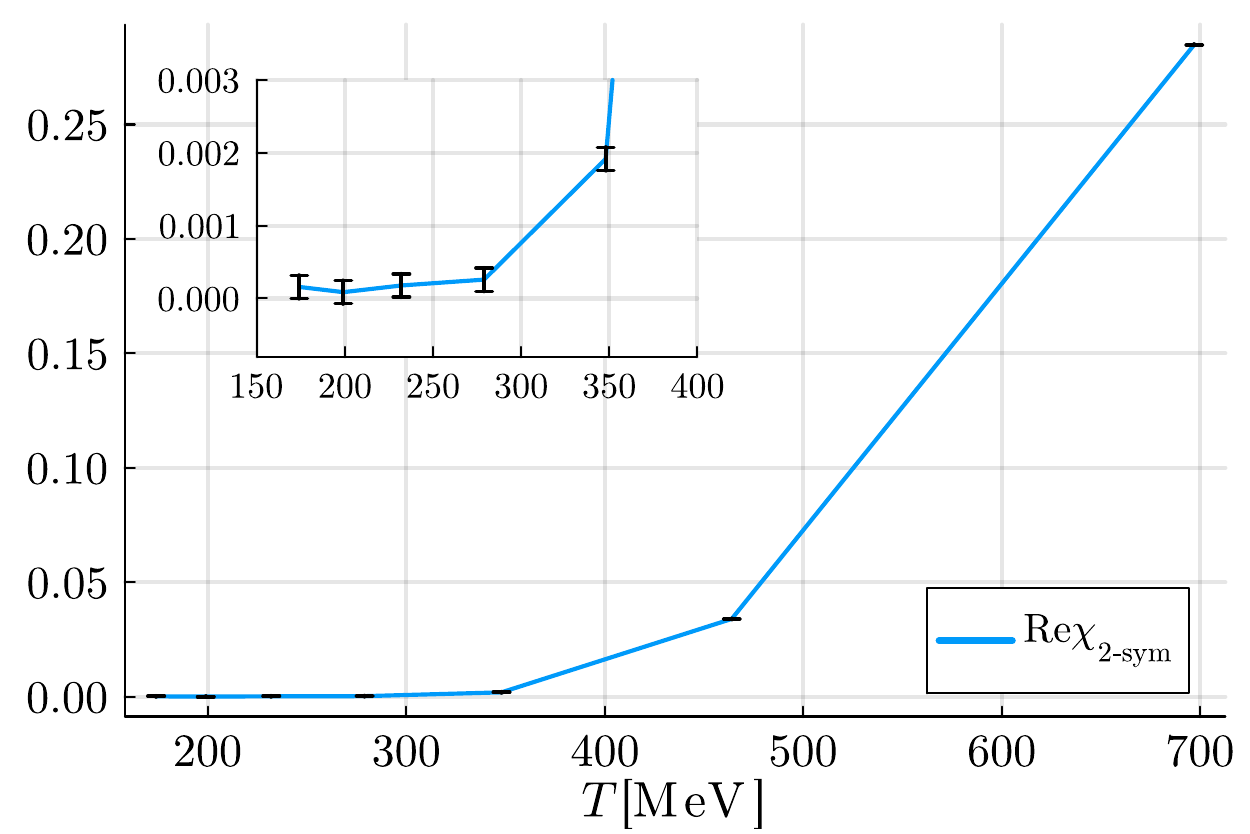}}
\scalebox{0.33}{
\includegraphics{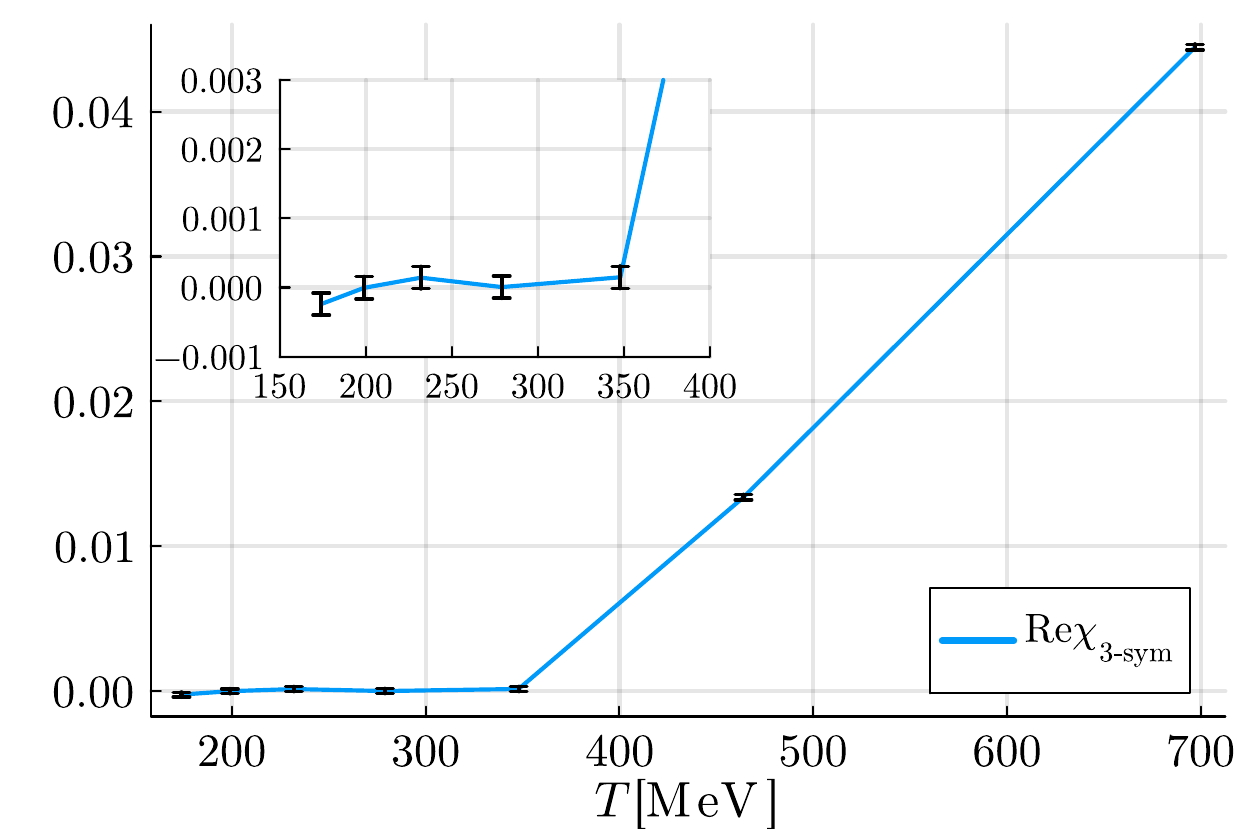}}
\end{center}
\caption{
The expectation values of characters vs. temperature for the fundamental, adjoint, rank-2 symmetric, and rank-3 symmetric representations, obtained from WHOT-QCD configurations. 
Note that the expectation values are real. 
}\label{fig:T-vs-character}
\end{figure}

\begin{figure}[htbp]
\begin{center}
\scalebox{0.4}{
\includegraphics{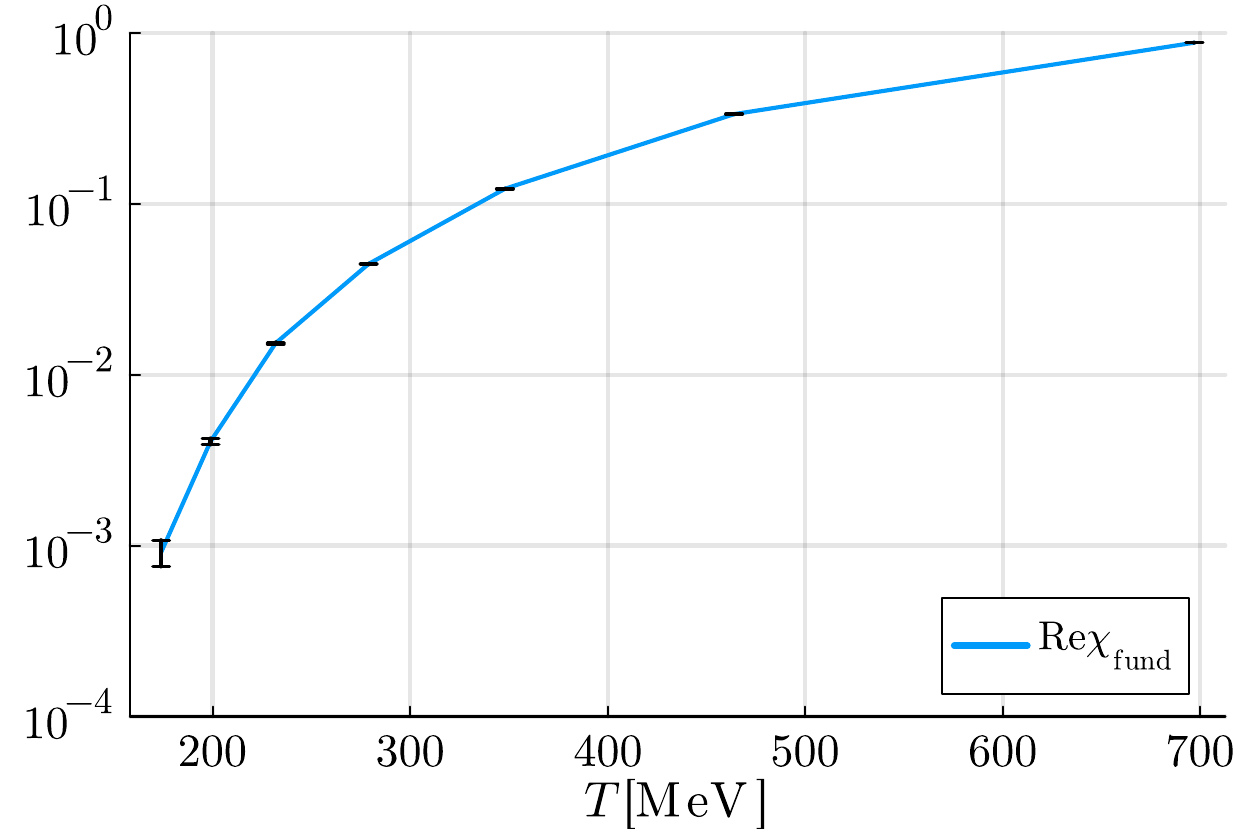}}
\end{center}
\caption{The log plot of the expectation value of the Polyakov loop in fundamental representation.
%$\chi_{\rm fundamental}$. 
It is very close to zero at the lowest temperature in the set of configurations (174~MeV). 
}\label{fig:T-vs-fund_character}
\end{figure}

Fig.~\ref{fig:T-vs-character} and Fig.~\ref{fig:T-vs-fund_character} show the expectation values of the Polyakov loops in several nontrivial representations.~\footnote{Note that the expectation values of Polyakov loops are real in the absence of the chemical potential, since $\rho(P)=\rho(P^\dagger)$. } These plots show, firstly, that the Polyakov loops do disappear at $T\lesssim 174$ MeV. In particular, Fig. \ref{fig:T-vs-fund_character} shows that 
the Polyakov loop in the fundamental representation is suppressed exponentially in the low-temperature regime.~\footnote{We cannot exclude the possibility of exponentially small deviation from the Haar-randomness at nonzero temperatures which is consistent with zero with our numerical precision. This is indeed the case for large-$N$ QCD on a small three-sphere~\cite{Hanada:2019kue}. See also the discussion section.}
Secondly, the expectation values of the Polyakov loops in higher representations become nonzero at different temperatures, $T\gtrsim 348$~MeV. 

The simplest possibility consistent with our observations is that there are three phases: (i) $T<T_1\sim 174$ MeV where the Polyakov line is governed by the Haar-random distribution, (ii) $T_1 < T < T_2 \sim 348  \rm{MeV}$ where the fundamental Polyakov loop is non-zero but the Polyakov loops associated with higher representations vanish, and (iii) $T_2 < T$ where Polyakov loop in all representations are non-zero. 

It is natural to interpret the nonzero values of the Polyakov loops in nontrivial representations 
as indicating that the degrees of freedom associated with the representations are deconfined. As such, we shall call the regimes, 
%(i), (ii), and (iii), 
(i) the completely-confined, (ii) partially-deconfined (or equivalently, partially-confined), and (iii) completely-deconfined phases, respectively. The identified phases are shown in Table~\ref{table:WHOT}.

\begin{figure}[hbtp]
\begin{minipage}{5cm}
\centering
\scalebox{0.35}{\includegraphics{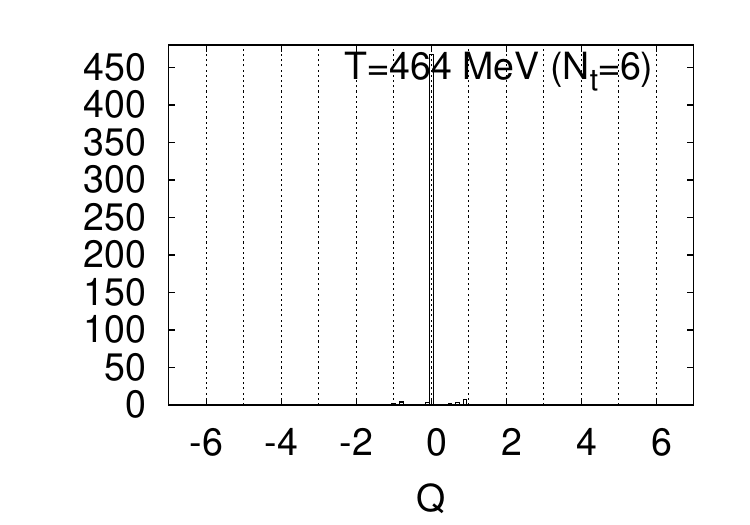}}
\end{minipage}
\begin{minipage}{5cm}
\centering
\scalebox{0.35}{\includegraphics{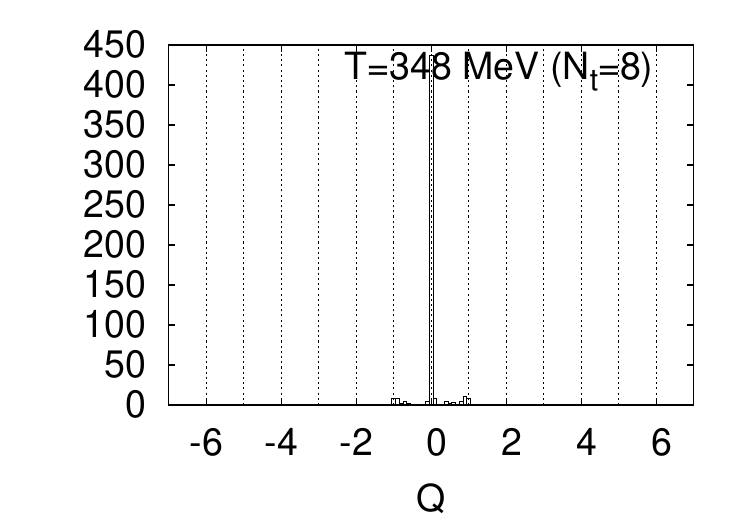}}
\end{minipage}
\begin{minipage}{5cm}
\centering
\scalebox{0.35}{\includegraphics{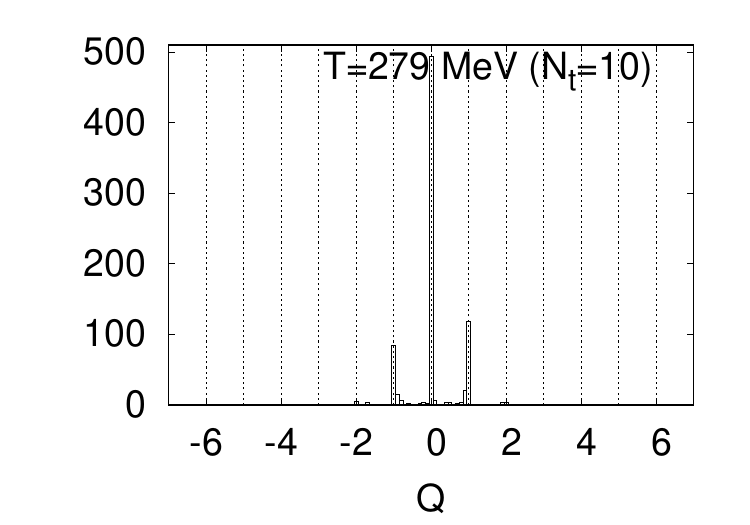}}
\end{minipage}
\begin{minipage}{5cm}
\centering
\scalebox{0.35}{\includegraphics{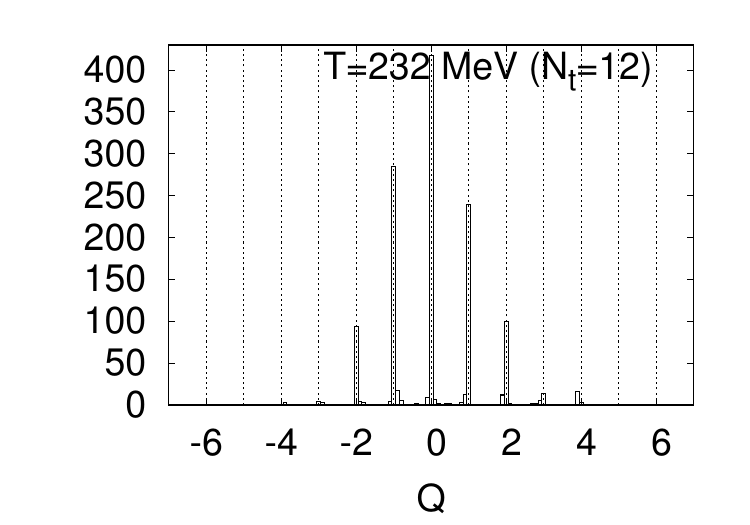}}
\end{minipage}
\begin{minipage}{6cm}
\centering
\scalebox{0.35}{\includegraphics{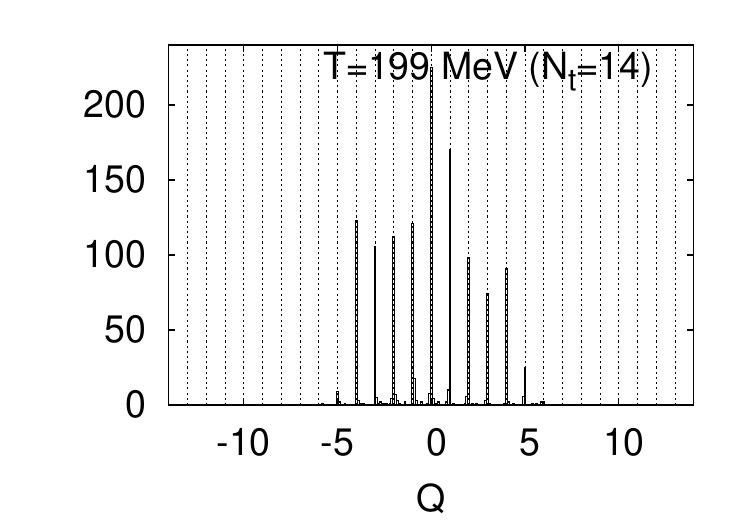}}
\end{minipage}
\begin{minipage}{5cm}
\centering
\scalebox{0.35}{\includegraphics{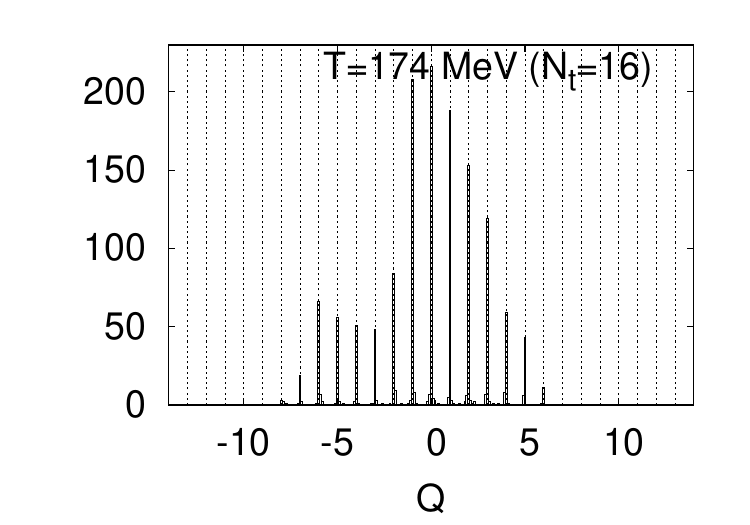}}
\end{minipage}
%\scalebox{0.38}{\includegraphics{Q_histogram_t8ov16_clover_imp_2856_51_1220.pdf}}
\caption{Histogram of the topological charge provided by the authors of Ref.~\cite{Taniguchi:2016tjc}. 
}\label{fig:topological-charge-WHOT-all-temperature}
\end{figure}

Remarkably, further support for this identification of the phases is obtained by studying the condensation of instantons.
Namely, we find that the instanton condensation occurs for temperature regimes corresponding to what
we identify as the completely-confined and partially-confined phases, whereas the instanton does not
condense in the completely-deconfined phase.
The quantity we use to detect the instanton condensation is the topological charge of each lattice configuration $Q$ computed by the WHOT-QCD collaboration~\cite{Taniguchi:2016tjc}. 
%Because lattice configurations are sensitive to the ultraviolet cutoff, they are smeared by using the gradient flow~\cite{Luscher:2010iy}. 
Because the topological charge is sensitive to the ultraviolet cutoff, it should be evaluated from a smeared lattice configuration, for example, by using the gradient flow~\cite{Luscher:2010iy}. 
After smearing, each configuration returns an integer value, or more precisely speaking, the histogram peaks at integer values. 
Fig.~\ref{fig:topological-charge-WHOT-all-temperature} shows the distributions of the topological charge at various temperatures.
At $T\le 279$~MeV, we clearly see multiple peaks including the ones at $Q\neq 0$ that signal the instanton condensation. (That the peaks gradually become lower and eventually disappear can be understood as the finite-volume effect.) The condensation melts as the temperature goes up. At $T=348$~MeV, we observe one sharp peak at $Q=0$ and almost vanishing peaks at $Q=\pm 1$. The instantons cease to condense around this temperature.

%%%%%%%%%%%%
%%%%%%%%%%%%
\section{Comparison to the large-$N$ partial deconfinement}\label{sec:comparison_with_large_N}
\hspace{0.51cm}
%%%%%%%%%%%%
%%%%%%%%%%%%
The observations in the previous section fit nicely within the framework of partial deconfinement~\cite{Hanada:2016pwv,Berenstein:2018lrm,Hanada:2018zxn,Hanada:2019czd} in large-$N$ theories. The close connection between the large-$N$ partial deconfinement and the behavior of QCD justifies the use of the terms we used to designate the phases: completely-confined, partially-deconfined, and  
completely-deconfined. 

To understand the meaning of partial deconfinement and Polyakov loop, the amount of gauge redundancy plays an important role~\cite{Hanada:2020uvt}. This is explained in Appendix~\ref{sec:meaning-Polyakov-large-N}. See also the companion paper~\cite{Hanada:2023rlk} which explains more details. 

In seminal papers~\cite{Aharony:2003sx, Sundborg:1999ue}, it was pointed out that the confinement/deconfinement transition consists of two phase transitions based on the weak coupling analysis. A more explicit understanding of the physical interpretation and the mechanism of the emergence of the intermediate phase was developed in a series of papers~\cite{Hanada:2018zxn,Hanada:2019czd,Hanada:2019kue,Hanada:2020uvt,Hanada:2023rlk}. Specifically, this phase was identified as the coexistence of confined and deconfined degrees of freedom in the space of colors (internal space) rather than in the usual coordinate space.

We can summarize the connection between the large-$N$ partial deconfinement and the new perspectives discussed above as follows.
\begin{enumerate}
    \item 
    The completely-confined phase is governed by the Haar-random distribution of the Polyakov lines~\cite{Hanada:2020uvt}. This is common to both large-$N$ theories and finite-$N$ theories including QCD.
    
\item The transition from the partially-deconfined phase to the completely-deconfined phase is identified with the Gross-Witten-Wadia (GWW) transition in the large-$N$ theory~\cite{Gross:1980he,Wadia:2012fr}. The GWW transition can be captured by using the expectation values of the multiply-wound Polyakov loops $u_n = \langle \tr P^n \rangle$. In particular, after the GWW transition, all $u_n$ become non-zero.~\footnote{More generally, $u_n$ with large $n$ may take nonzero but exponentially suppressed values below the GWW point, as is the case for large-$N$ QCD on a small three-sphere~\cite{Schnitzer:2004qt,Hanada:2019kue}.}
For the QCD, we find that the transition from the partially-deconfined phase to the completely-deconfined phase is associated with the onsets of the Polyakov loops in the higher representations.  The Polyakov loops in the higher representations and $u_n$'s (with $n \geq 1$) play analogous roles and some of them are directly related. In Appendix \ref{sec:character}, we give examples of direct relations between $u_n$ and Polyakov loops in higher representations.

\item As discussed and observed in Refs.~\cite{Hanada:2019kue,Hanada:2021ksu}, it is natural to expect that the chiral symmetry breaking takes place at the GWW point when quarks are massless. (Intuitively, quarks in the confined sector should form a chiral condensate; otherwise, the 't Hooft anomaly would not be preserved.)
Since the instantons are intimately connected with the chiral symmetry, it is natural to expect the behavior of instantons to change across 
the GWW transition.~\footnote{
It was shown in Ref.~\cite{Buividovich:2015oju} that the
GWW transition of 2D lattice Yang-Mills theory can be directly understood as arising from the change of the saddle points contributing to the path integral including the contribution of the instantons. }
Since in QCD the quark mass breaks the chiral symmetry explicitly, the instanton condensation is a natural probe for the finite-$N$ analog of the GWW transition.
Indeed, we find that the phase structure suggested by instantons is consistent with that obtained from the Polyakov loops. 

\end{enumerate}

%%%%%%%%%%%%%%%%%%%%%%
%%%%%%%%%%%%%%%%%%%%%%
\section{Discussion}
%%%%%%%%%%%%%%%%%%%%%%
%%%%%%%%%%%%%%%%%%%%%%
We have discussed the thermal phase transitions on 4d QCD, i.e., SU(3) gauge group and dynamical quarks, as a natural extension of partial deconfinement in the large-$N$ gauge theories.
Employing the gauge configuration generated by the WHOT-QCD collaboration, we found numerically that the Polyakov loops in various representations start to have nonzero expectation values at different temperatures.\footnote{
Partial deconfinement is a generic property of gauge theories insensitive to the details of the theories such as gauge group or matter content, at least at large $N$. 
Therefore, it is natural to expect other finite-$N$ theories to exhibit thermal phases that can be characterized like what we have done for QCD.
} 
Our main point is that these should play an important role in characterizing the phases of QCD at finite temperatures.

While we found that some expectation values are consistent with zero at the current numerical precision, their true behavior may be $\ev{\chi_r(P)}\sim e^{-m_r/T}$ where $m_r$ is the mass gap from the vacuum (at zero temperature) for the degrees of freedom associated with the representation $r$.~\footnote{In the case of large-$N$ QCD on a small three-sphere, this is indeed the case~\cite{Schnitzer:2004qt,Hanada:2019kue}.}
This is so in particular because it is not forbidden by center symmetry.
Our interpretation is that deviation from this scaling detects the deconfinement of the modes in the corresponding representation.

In this letter, we focused on the bare, rather than the renormalized, Polyakov loops to characterize the phases of thermal QCD.
It is crucial that we study the theory defined at the identical UV cutoff for various temperatures;
the configurations use fixed lattice spacing and the temperature is controlled by changing the number of lattice points along the Euclidean time direction.  As such, renormalization is not necessary for our characterization.

However, we should emphasize that this is based on the viewpoint of a cutoff field theory. To compare configurations with different cutoff scales and to take the continuum limit, i.e., to confirm that our characterization of phases works properly in QFTs, it is necessary to rephrase our analysis in terms of renormalized Polyakov loops. If the expectation values of the Polyakov loops are multiplicatively renormalized without mixing, we may expect the characterization to be unchanged even if we use the renormalized Polyakov loops instead of the bare Polyakov loops. 
There are various prescriptions for defining renormalized Polyakov loops. 
See, for example, \cite{Kaczmarek:2002mc,Dumitru:2003hp,Gupta:2006qm,Gupta:2007ax,Mykkanen:2012ri,Petreczky:2015yta} and references therein.
However, it is nontrivial whether one can apply appropriate renormalization schemes that satisfy the multiplicative renormalizability and also preserve the ``exponential smallness’’ of the expectation values in the (completely and partially) confined phases. This is an important problem open for questions and it is worthwhile to explore as a future direction.

To overcome the issue of renormalization, it would be useful to make use of other quantities to distinguish phases. In this letter, we focused on the one-point function of the Polyakov loops.  It is important to consider other observables, such as multi-point correlation functions, from the point of view of the Haar-randomness and the deviation from it. See Ref.~\cite{Bergner:2023rpw} for development along this line, deriving the so-called Casimir scaling of Polyakov loops in various representations from this point of view. The smearing and gradient flow might be useful in this context.

Our conjectured phase structure of thermal QCD is as follows;
The first `transition' at $T_1$ may well be a crossover. For example, the Polyakov loops are allowed to have small nonzero values that become exponentially small when the sizes of the representations are increased.
It would be more natural to expect a transition with non-analyticity at $T_2$, given the connection to the condensation of instantons. 
This can be the case even if the Polyakov loops in large representations are not exactly zero at $T<T_2$;
a natural possibility would be that there is a transition between the exponential decay with respect to the dimension of the representation at $T<T_2$ and the power-law decay at $T>T_2$.
Note that the possible presence of the intermediate phase in the region $T_\mathrm{c} \le T \lesssim 3T_\mathrm{c}$ (where $T_\mathrm{c}$ denotes the usual QCD critical temperature) has been discussed from various perspectives (see, e.g.,~\cite{Asakawa:1995zu,Glozman:2022zpy,Cohen:2023hbq}).

It is clearly important to further verify the new perspective we advocated in this letter. In particular, studying the Polyakov loop in larger representations and its behavior when departing from Haar-randomness, identifying the transition temperature and the order of the phase transitions, and establishing suitable renormalization schemes are valuable future problems. Practically, SU(3) QCD with finite quark mass and pure Yang-Mills are the most tractable targets because many sets of lattice configurations are available for SU(3) QCD and the simulation of pure Yang-Mills is not costly.

An obstacle to generalizing partial deconfinement to finite $N$ had been the meaning of the size of the deconfined sector $M$ ($0\le M \le N$). Even in the large-$N$ limit, $M$ is not literally an integer. It could have an uncertainty of order $N^0$ which is negligible at large $N$. Admittedly, such an ambiguity makes the use of ``$M$" very subtle at $N=3$. To circumvent this issue, we avoided the use of $M$ and relied on characters (the Polyakov loops in various representations).
The use of the character also has the advantage of being manifestly gauge invariant. Although in large $N$-theory the use of the parameter $M$ is shown to have gauge invariant meaning, it may be worthwhile to revisit the analysis of the partial deconfinement in the large-$N$ theory using the character expansion. Character expansion played an important role in the large-$N$ theory, for example in Refs.~\cite{Gross:1993hu, Gross:1993yt}.  A recent work~\cite{Berenstein:2023srv} employs the character expansion to study the deconfinement transition
 in large-$N$ theory.

In this letter, we have explored the finite-$N$ counterpart of the large-$N$ partial deconfinement. 
The original motivation for partial deconfinement~\cite{Hanada:2016pwv} was to study black hole geometry via holography.
Since $1/N$ corrections should play crucial roles in black hole physics, we hope that our work would be useful in obtaining important intuition into quantum gravitational phenomena such as black hole evaporation from the QFT side. 
%%%%%%%%%%%%%%%%%%%%%%
%%%%%%%%%%%%%%%%%%%%%%
\begin{center}
\Large{\textbf{Acknowledgement}}
\end{center}
%%%%%%%%%%%%%%%%%%%%%%
%%%%%%%%%%%%%%%%%%%%%%
We would like to thank the members of the WHOT-QCD collaboration, including Shinji Ejiri, Kazuyuki Kanaya, Masakiyo Kitazawa, and Takashi Umeda, for providing us with their lattice configurations and many plots and having stimulating discussions with us. 
The analysis of topological charge in Ref.~\cite{Taniguchi:2016tjc} was led by Yusuke Taniguchi who passed away in 2022. Kazuyuki Kanaya collected the data and plots created by Yusuke Taniguchi for us. We deeply thank Yusuke Taniguchi and Kazuyuki Kanaya. 
%Shohei Ohtani's unbelievable performance gave the authors the energy to survive the hot and humid summer of 2023. 
We also thank Sinya Aoki, Hidenori Fukaya, Vaibhav Gautam, Yui Hayashi, Jack Holden, Seok Kim, Tamas Kovacs, Atsushi Nakamura, Marco Panero, Robert Pisarski, Enrico Rinaldi, Hideo Suganuma, Yuya Tanizaki, and Jacobus Verbaarschot for discussions and comments. 

This work was supported by the Japan Lattice Data Grid (JLDG) constructed over the SINET5 of NII and by the Center for Gravitational Physics and Quantum Information (CGPQI) at Yukawa
Institute for Theoretical Physics. 
M.~H. thanks for the STFC grants ST/R003599/1 and ST/X000656/1.
H.~O. is supported by a Grant-in-Aid for JSPS Fellows (Grant No.22KJ1662).
H.~S. is supported by the Japan Society for the Promotion of Science (JSPS) KAKENHI Grant number 21H05182.
H.~W. is supported by the Japan Society for the Promotion of Science (JSPS) KAKENHI Grant number 22H01218.

\appendix

%%%%%%%%%%%%
%%%%%%%%%%%%
\section{Polyakov loop and the amount of redundancy at $N=\infty$}\label{sec:meaning-Polyakov-large-N}
\hspace{0.51cm}
%%%%%%%%%%%%
%%%%%%%%%%%%
The purpose of this appendix is to give a short summary of the essential mechanism of the large-$N$ partial deconfinement.
In particular, we explain the relevance of the redundancy of states under gauge transformations and its relation to the Polyakov line.

That the amount of gauge redundancy has important consequences is not really new. In fact, it has been known for a century since the theoretical discovery of Bose-Einstein condensation~\cite{einstein1924quantentheorie}, although the connection to Polyakov line and confinement was pointed out only recently~\cite{Hanada:2020uvt}. 
To see how the Polyakov line and gauge redundancy are related, we consider three theories with increasing levels of complexity: $N$ indistinguishable bosons, SU($N$) Hermitian matrix model, and SU($N$) QCD. 
%%%%%%%%%%%%
%%%%%%%%%%%%
\subsubsection*{$N$ indistinguishable bosons and Bose-Einstein condensation}
\hspace{0.51cm}
%%%%%%%%%%%%
%%%%%%%%%%%%
To describe $N$ indistinguishable bosons in $\mathbb{R}^3$, we use coordinate operators $\hat{\vec{x}}_i=(\hat{x}_i, \hat{y}_i, \hat{z}_i)$ and momentum operators $\hat{\vec{p}}_i=(\hat{p}_{x,i}, \hat{p}_{y,i}, \hat{p}_{z,i})$, where $i=1,2,\cdots,N$ labels bosons. The Hamiltonian $\hat{H}$ is invariant under the permutation of the labels, e.g., $\hat{H}=\sum_{i=1}^N(\frac{1}{2m}\hat{\vec{p}}_i^2+\frac{m\omega^2}{2}\hat{\vec{x}}_i^2)$ in the weak-coupling limit. That the bosons are indistinguishable means the S$_N$ permutation symmetry is gauged. 

We can use the coordinate eigenstates $\ket{x}=\ket{\vec{x}_1,\cdots,\vec{x}_N}$ that satisfy $\hat{\vec{x}}_{i}\ket{x}=\vec{x}_{i}\ket{x}$ to describe quantum states. The coordinate eigenstates span the extended Hilbert space ${\cal H}_{\rm ext}$ that contains S$_N$-non-singlets:
\begin{align}
\mathcal{H}_{\rm ext}
=
\textrm{Span}\{
\ket{x}|x\in\mathbb{R}^{3N}
\}. 
\end{align}
For a permutation $\sigma\in\text{S}_N$, we define $\hat{\sigma}$ by $\hat{\sigma}\ket{\vec{x}_1,\cdots,\vec{x}_N}=\ket{\vec{x}_{\sigma(1)},\cdots,\vec{x}_{\sigma(N)}}$. From this, we can define the projection operator $\hat{\pi}=\frac{1}{N!}\sum_{\sigma\in\text{S}_N}\hat{\sigma}$ that maps $\mathcal{H}_{\rm ext}$ to the S$_N$-invariant Hilbert space $\mathcal{H}_{\rm inv}$. Canonical partition function at temperature $T$ can be written in two ways, in terms of $\mathcal{H}_{\rm ext}$ and $\mathcal{H}_{\rm inv}$:
\begin{align}
Z(T)
=
\mathrm{Tr}_{\mathcal{H}_{\rm inv}}(e^{-\hat{H}/T})
=
\mathrm{Tr}_{\mathcal{H}_{\rm ext}}(\hat{\pi}e^{-\hat{H}/T})
=
\frac{1}{N!}\sum_{\sigma\in\mathrm{S}_N}
\mathrm{Tr}_{\mathcal{H}_{\rm ext}}(\hat{\sigma}e^{-\hat{H}/T})\, . 
\label{N-boson-partition-function}
\end{align}

To see how states in $\mathcal{H}_{\rm inv}$ and $\mathcal{H}_{\rm ext}$ are related, let us consider the weak-coupling limit and consider a product of one-particle states, $\ket{\Phi}$, whose wave function is written as $\bra{x}\ket{\Phi}=\prod_{i=1}^N\phi_i(\vec{x}_i)$. The symmetric group S$_N$ acts on $\ket{\Phi}$ as $\bra{x}\hat{\sigma}\ket{\Phi}=\prod_{i=1}^N\phi_i(\vec{x}_{\sigma(i)})$. If all one-particle states are the same, i.e., $\phi_1=\cdots=\phi_N$, the product state is invariant under S$_N$. On the other hand, if $\phi_1,\cdots,\phi_M$ are all different but $\phi_{M+1}=\cdots=\phi_N$, then the product state is invariant under $\mathrm{S}_{N-M}\subset \mathrm{S}_{N}$. Such an unbroken symmetry acting on $\mathcal{H}_{\rm ext}$ leads to an enhancement factor in the partition function. Specifically,  when the overlap of different one-particle states can be neglected, we obtain
\begin{align}
\sum_{\sigma\in\mathrm{S}_N}
\bra{\Phi}\hat{\sigma}\ket{\Phi}
=
(N-M)!\, . 
\end{align}
This enhancement factor is responsible for the Bose-Einstein condensation. 
Many particles fall into the one-particle ground state assisted by this enhancement factor. 
From this, one can learn that \textit{less redundant states are preferred at low energy}. 

We will next discuss the SU($N$) Hermitian matrix model and will see that the permutation $\sigma$ is nothing but the Polyakov line. 
This implies that the typical Polyakov line in the path integral is determined in such a way that it leaves typical states dominating partition function invariant, i.e., $\hat{\sigma}\ket{\rm typical}=\ket{\rm typical}$. 
We will elaborate on this statement shortly. 

%%%%%%%%%%%%
%%%%%%%%%%%%
\subsubsection*{SU($N$) Hermitian matrix model and color confinement}
\hspace{0.51cm}
%%%%%%%%%%%%
%%%%%%%%%%%%
Let us consider a bosonic SU($N$) gauged $D$-matrix model ($D\ge 2$) with the Hamiltonian
\begin{align}
\hat{H}=
{\rm tr}\left(
\frac{1}{2}\sum_{I=1}^D\hat{P}_I^2
+
V(\hat{X})
\right)\, , 
\end{align}
where ${\rm tr}$ means the trace as $N\times N$ matrix and 
$V(\hat{X})$ is a potential term such as $V(\hat{X})=-\frac{g^2}{4}[\hat{X}_I,\hat{X}_J]^2$.  
Each $\hat{X}_I$ has $N^2$ components $\hat{X}_{I,ij}$, where $i,j=1,2,\cdots,N$, that satisfy the Hermiticity condition $(\hat{X}_{I,ij})^\dagger=\hat{X}_{I,ji}$. 
We do not impose the traceless condition. 
The operator $\hat{P}_{I,ij}$ is the conjugate momentum of $\hat{X}_{I,ji}$. 
They satisfy the canonical commutation relation 
$[\hat{X}_{I,ij},\hat{P}_{J,kl}]=i\delta_{IJ}\delta_{il}\delta_{jk}$.

The Hamiltonian is invariant under the adjoint action of SU($N$) defined by
\begin{align}
\hat{X}_{I,ij}
\to
(U\hat{X}_{I}U^{-1})_{ij}
=
\sum_{k,l}U_{ij}\hat{X}_{I,kl}U^{-1}_{lj}, 
\qquad
\hat{P}_{I,ij}
\to
(U\hat{P}_{I}U^{-1})_{ij}
=
\sum_{k,l}U_{ik}\hat{P}_{I,kl}U^{-1}_{lj}. 
\end{align}

We can use the extended Hilbert space with SU($N$) non-singlet states, ${\cal H}_{\rm ext}$. 
The extended space is spanned by the coordinate eigenstates $\ket{X}$ that satisfy $\hat{X}_{I,ij}\ket{X}=X_{I,ij}\ket{X}$:
\begin{align}
\mathcal{H}_{\rm ext}
=
\textrm{Span}\{
\ket{X}|X\in\mathbb{R}^{DN^2}
\}. 
\end{align}
Note that $X$ consists of $DN^2$ real numbers $X_{I=1,\cdots,D}^{\alpha=1,\cdots,N^2}$. 
Gauge transformation acts on $\mathcal{H}_{\rm ext}$ as
$\ket{X}
\to
\ket{U^{-1} XU}$.
By using the SU($N$)-invariant Hilbert space ${\cal H}_{\rm inv}$, the canonical partition function at temperature $T$ can be written as~\cite{Hanada:2020uvt} 
\begin{align}
Z(T)
=
{\rm Tr}_{\mathcal{H}_{\rm inv}}\left(
e^{-\hat{H}/T}
\right)
=
\mathrm{Tr}_{\mathcal{H}_{\rm ext}}(\hat{\pi}e^{-\hat{H}/T})
=
\frac{1}{\textrm{Vol}G}\int_{G} dg
\mathrm{Tr}_{\mathcal{H}_{\rm ext}}(\hat{g}e^{-\hat{H}/T})\, , 
\label{MM-partition-function}
\end{align}
where $G=\mathrm{SU}(N)$ and
$\hat{\pi}
\equiv
\frac{1}{\textrm{vol}G}\int_G dg\,
\hat{g}$
is a projection operator from ${\cal H}_{\rm ext}$ to ${\cal H}_{\rm inv}$. 
\begin{figure}[htbp]
\begin{center}
\scalebox{0.3}{
\includegraphics{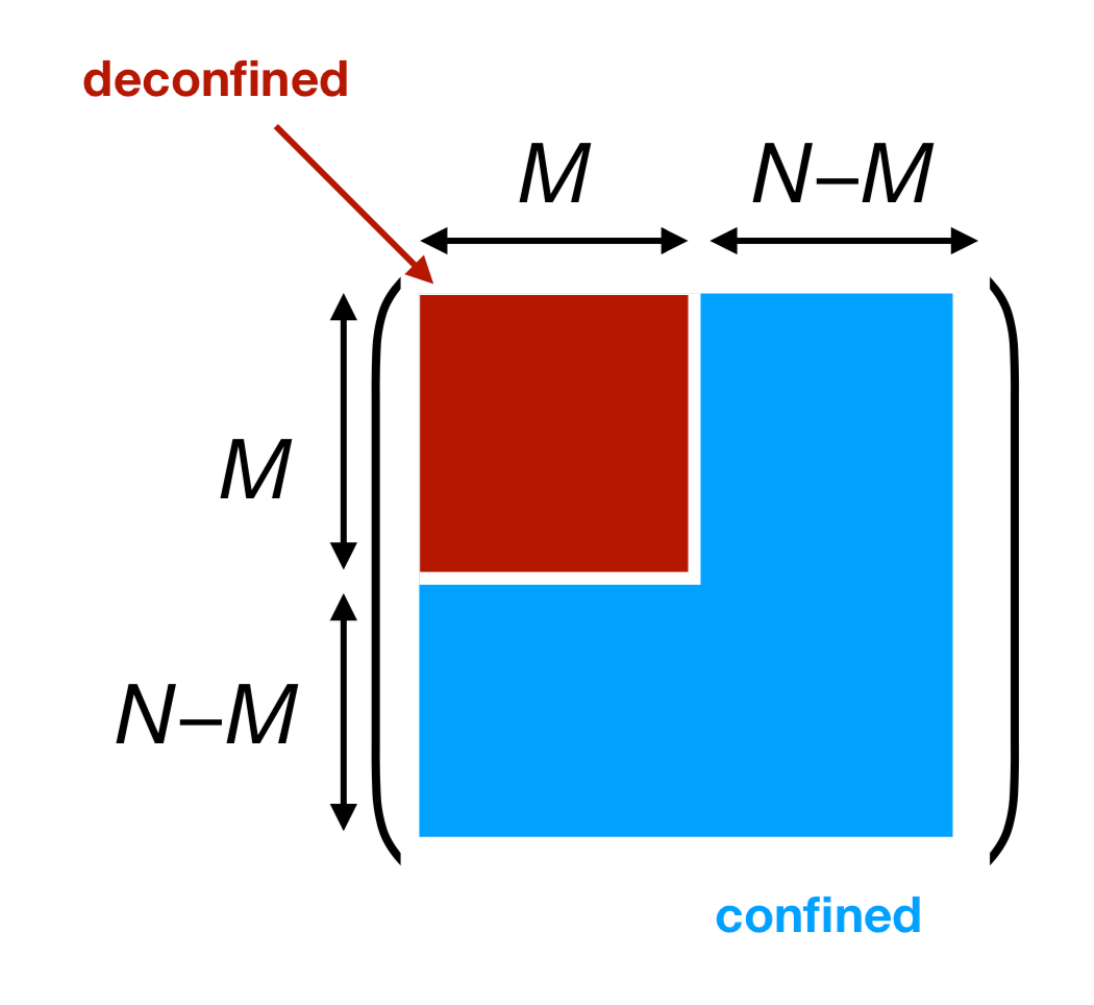}}
\end{center}
\caption{Partial deconfinement in the matrix model. The $M\times M$-block shown in red is deconfined. This figure is taken from Ref.~\cite{Hanada:2019czd}. 
}\label{fig:matrix}
\end{figure}

The standard technique to rewrite the Hamiltonian formulation to the path-integral formulation tells us that $g\in G$ is the Polyakov line~\cite{Hanada:2020uvt}. By comparing \eqref{MM-partition-function} with \eqref{N-boson-partition-function}, we see that $\sigma\in\mathrm{S}_N$ corresponds to the Polyakov line. The counterpart of the Bose-Einstein condensation is confinement~\cite{Hanada:2020uvt}. 
In general, SU($M$) subgroup of SU($N$) can be deconfined~\cite{Hanada:2016pwv,Berenstein:2018lrm,Hanada:2018zxn,Hanada:2019czd}, leaving $\mathrm{SU}(N-M)\subset\mathrm{SU}(N)$ as a symmetry in $\mathcal{H}_{\rm ext}$ that leads to the enhancement factor $\mathrm{Vol}(\mathrm{SU}(N-M))\sim e^{(N-M)^2}$~\cite{Hanada:2020uvt}. 
The value of $M$ depends on the energy $E$ in a nontrivial manner. 
%%%%%%%%%%%%
%%%%%%%%%%%%
\subsubsection*{Distribution of Polyakov line phases in the large-$N$ limit}
\hspace{0.51cm}
%%%%%%%%%%%%
%%%%%%%%%%%%
Let us further focus on typical states dominating thermodynamics that are invariant under $\textrm{SU}(N-M)\subset\textrm{SU}(N)$. 
We can fix the gauge in such a way that the SU($M$)-deconfined sector is the upper-left $M\times M$-block as in Fig.~\ref{fig:matrix}. This choice of embedding of SU($M$) into SU($N$) fixes SU($N$) down to $\mathrm{SU}(M)\times\mathrm{SU}(N-M)$. Then, the Polyakov line $\mathcal{P}$ takes the following form:
\begin{align}
    \mathcal{P}=
    \left(
    \begin{array}{cc}
    \mathcal{P}_{\rm dec} & 0\\
    0 & \mathcal{P}_{\rm con}
    \end{array}
    \right)\, .
    \label{eq:gauge-fixing-Polyakov}
\end{align}
The eigenvalues of $\mathcal{P}_{\rm dec}$ and $\mathcal{P}_{\rm con}$ give the phases  $\theta_1,\cdots,\theta_M$ and $\theta_{M+1},\cdots,\theta_N$, respectively. 
From them, we can determine the distribution of the phases $\rho_{\rm dec}(\theta)$ and $\rho_{\rm con}(\theta)$. The latter is constant,
\begin{align}
    \rho_{\rm con}(\theta)=\frac{1}{2\pi}\, . 
\end{align}
This is because $\mathcal{P}_{\rm con}$ can be any element of SU($N-M$) (specifically, $\mathcal{P}_{\rm con}$ dominating path integral is distributed following the Haar measure) and the generic phase distribution in this part is uniform in the limit of $N-M\to\infty$. 
The former is not uniform and its smallest value is zero. 
The full distribution is~\cite{Hanada:2020uvt,Hanada:2018zxn,Hanada:2019czd} 
\begin{align}
    \rho(\theta)
    =
    \frac{1}{2\pi}\cdot\left(1-\frac{M}{N}\right)
    +
    \frac{M}{N}\cdot\rho_{\rm dec}(\theta)\, . 
    \label{phase:full=con+dec}
\end{align}
Therefore, the constant offset provides us with a way to measure the amount of gauge redundancy:
\begin{align}
\mathrm{\ Constant\ offset}
=
\frac{1}{2\pi}\left(1-\frac{M}{N}\right)\, . 
\end{align}
This relation holds for the system of indistinguishable bosons as well. 
%%%%%%%%%%%%%%%%%%%%%%
%%%%%%%%%%%%%%%%%%%%%%
\subsubsection*{SU($N$) QCD}
%%%%%%%%%%%%%%%%%%%%%%
%%%%%%%%%%%%%%%%%%%%%%
Essentially the same mechanism works for large-$N$ QCD. Although we leave the detail to the companion paper~\cite{Hanada:2023rlk}, the essence is easy to understand:
If the fields are slowly varying up to gauge transformations (which must be the case at low energy), in a small enough spatial volume we can reduce the theory to a matrix model (i.e., a theory without spatial extent), and then, the same mechanism analogous to BEC works. 

%%%%%%%%%%%%%%%%%%%%%%%%%%%%%%%%
%%%%%%%%%%%%%%%%%%%%%%%%%%%%%%%%
\section{SU($N$) Haar-random distribution}\label{sec:Haar-random-distribution}
%%%%%%%%%%%%%%%%%%%%%%%%%%%%%%%%
%%%%%%%%%%%%%%%%%%%%%%%%%%%%%%%%
In this appendix, we will give an elementary proof, for $N=3$, of the analytic formula for the Haar-random distribution of the phases in SU($N$),  
\begin{align}
\rho_{\rm Haar}(\theta)=\frac{1}{2\pi}\left(1-(-1)^N\cdot\frac{2}{N}\cos(N\theta)\right).
\label{eq:appendix:Haar-random}
\end{align}
See Ref.~\cite{Hanada:2023rlk} for other values of $N$, and Ref.~\cite{Nishigaki:2024phx} for a rigorous proof of this formula for arbitrary $N$ and its natural generalizations for other gauge groups.

%\textcolor{red}{In what follows, we show a primitive proof for $N=3$.}
We start with
\begin{equation}
    \qty[\dd U] 
    =
    \rho(\theta_1,\cdots,\theta_N)
    \prod_{j=1}^N \dd \theta_j\, ,
\end{equation}
where
\begin{align}
    \rho(\theta_1,\cdots,\theta_N)
    =
    C \prod_{j<k} \sin^2\left(\frac{\theta_j - \theta_k}{2}\right)
    \times
    \sum_{n=-\infty}^{\infty}
    \delta\left(
        \sum_{j=1}^N \theta_j -2\pi n
    \right)\, . 
    \label{eq:SU(N)_Haar_measure}
\end{align}
Here, $C$ is the normalization factor and $\prod_{j<k} \sin^2\left(\frac{\theta_j - \theta_k}{2}\right)$ is the Vandermonde determinant. 
Our task is to integrate out $\theta_2,\cdots,\theta_N$ and estimate
\begin{align}
    \rho(\theta_1)
    =
    \int\dd\theta_2\cdots\int\dd\theta_N\ 
    \rho(\theta_1,\theta_2,\cdots,\theta_N)\, . 
\end{align}
%%%%%%%%%%%%%%%%%%%%%%%%%%%
%\subsubsection*{$N=3$ case}
%%%%%%%%%%%%%%%%%%%%%%%%%%%
For $N=3$, we use  $\theta_3 = -(\theta_1 + \theta_2)$ up to the shift by $2\pi n$.
By substituting this into \eqref{eq:SU(N)_Haar_measure}, we obtain 
\begin{align}
    \rho(\theta_1,\theta_2)
    &=
    C
    \sin^2\left(\frac{\theta_1-\theta_2}{2}\right)
    \sin^2\left(\frac{\theta_1 + 2\theta_2}{2}\right)
    \sin^2\left(\frac{\theta_2 + 2\theta_1}{2}\right)
    \nonumber\\
    &=
    \frac{C}{8}
    \left(1-\cos(\theta_1-\theta_2)\right)
    \left(1-\cos(\theta_1+2\theta_2)\right)
    \left(1-\cos(2\theta_1+\theta_2)\right)
    \nonumber\\
    &=    
    \frac{C}{8}
    \left(
    \frac{3}{4}
    +
    \frac{1}{2}\cos(3\theta_1)
    +
    \cdots
    \right)\, ,          
\end{align}
where $\cdots$ are terms like $\cos(\theta_1-\theta_2)$ or $\cos(\theta_1+2\theta_2)$ which disappear after $\theta_2$ is integrated. By integrating $\theta_2$, we obtain 
\begin{equation}
    \rho(\theta_1) 
    =
    \int_{-\pi}^\pi \dd \theta_2\, \rho(\theta_1,\theta_2)
    \propto
    1 + \frac{2}{3}\cos(3\theta_1)\, . 
\end{equation}
We can fix the overall factor from  $\int_{-\pi}^\pi \dd \theta_1\, \rho(\theta_1)=1$. 
%%%%%%%%%%%%%%%%%%%%%%%%%%%%%
\section{Characters of SU(3) group}\label{sec:character}
%%%%%%%%%%%%%%%%%%%%%%%%%%%%%
The character of $G$ is defined by the trace of the representation matrix $R_r$ in the representation $r$ as
\begin{equation}
    \chi_r(g) := \tr R_r(g).
\end{equation}
For the irreducible representations, the orthonormal condition,
\begin{equation}
    \frac{1}{\mathrm{Vol}(G)} \int_G \mathrm{d}g\, \chi_r(g)\left(\chi_s(g)\right)^\ast = \delta_{rs},
    \label{eq:orthogonality}
\end{equation}
is satisfied.

When we consider $G=$ SU(3), we can identify $g$ and its representation in the fundamental representation: $g$ is $3\times 3$ unitary matrix with a determinant equal to 1. 
In our context $g$ is identified with the Polyakov line $P$. 
The SU(3) characters are functions only with respect to the eigenvalues of $g$, denoted by $\lambda_j$ $ (j=1,2,3)$. 
Namely, 
\begin{equation}
    \chi_r(g) = \chi_r(\{\lambda\}).
\end{equation}
Note that $\lambda_j = e^{i\theta_j}$ in terms of the Polyakov line phase $\theta_j$, and $\lambda_1\lambda_2\lambda_3=1$ due to the condition $\det g = 1$.
%%%%%%%%%%%%%%%%%%%%%%%%%%%%%%%%%%%%%
%%%%%%%%%%%%%%%%%%%%%%%%%%%%%%%%%%%%%
\subsection{List of characters for irreducible representations}
%%%%%%%%%%%%%%%%%%%%%%%%%%%%%%%%%%%%%
%%%%%%%%%%%%%%%%%%%%%%%%%%%%%%%%%%%%%
The character is a symmetric polynomial in $\lambda$’s. 
As is well-known, irreducible representations of $\rm{SU}(3)$ can be labelled by Young diagram with two rows.
In the following, the notation $(n,m)$ represents a Young tableau that has $n$ and $m$ boxes in the first and second rows, respectively.

\begin{description}
%%%%%%%%%%%%%%%%%%%%%%%
\item[(0,0)=1: trivial]
%%%%%%%%%%%%%%%%%%%%%%%
%Trivial representation is $R_{(0,0)}(g)=1$. The character is
\begin{equation}
    \chi_{\rm trivial}
    \equiv
    \chi_{(0,0)} = 1\, . 
\end{equation}

%%%%%%%%%%%%%%%%%%%%%%%%%%%
\item[(1,0)=3: fundamental]
%%%%%%%%%%%%%%%%%%%%%%%%%%%
%Fundamental representation is $3\times 3$ special unitary matrix $g$ itself: $R_{(1,0)}(g)=g$. Hence the character is $\mathrm{Tr}g$, which is the sum of three eigenvalues $\lambda_1$, $\lambda_2$, and $\lambda_3$:
\begin{equation}
    \chi_{\rm fund.}
    \equiv
    \chi_{(1,0)}
    =
    \lambda_1 + \lambda_2 + \lambda_3\, . 
\end{equation}
%%%%%%%%%%%%%%%%%%%%%%%%%%%%%%%%%%%%%%%%%
\item[(1,1)=$\bar{\bf 3}$: rank-two anti-symmetric, or anti-fundamental]
%%%%%%%%%%%%%%%%%%%%%%%%%%%%%%%%%%%%%%%%%
\begin{equation}
    \chi_{\rm antisym.}
    \equiv
    \chi_{(1,1)}
    =
    \lambda_1\lambda_2 + \lambda_2\lambda_3 + \lambda_3\lambda_1
    =
    \lambda_1^{-1}
    +
    \lambda_2^{-1}
    +
    \lambda_3^{-1}
    =
    (\chi_{\rm fund.})^\ast\, . 
\end{equation}
%%%%%%%%%%%%%%%%%%%%%%
\item[(2,0)=6: rank-two symmetric]
%%%%%%%%%%%%%%%%%%%%%%
\begin{equation}
    \chi_{(2,0)}
    =
    \lambda_1^2 + \lambda_2^2 + \lambda_3^2 + 
    \lambda_1\lambda_2 + \lambda_2\lambda_3 + \lambda_3\lambda_1\, . 
\end{equation}
%%%%%%%%%%%%%%%%%%%%%%%
\item[(2,1)=8: adjoint]
%%%%%%%%%%%%%%%%%%%%%%%

\begin{equation}
    \chi_{\rm adj.}
    \equiv
    \chi_{(2,1)}
    =
    2 + 
    \sum_{a\neq b}
    \lambda_a\lambda_b^{-1}\, . 
\end{equation}

%%%%%%%%%%%%%%%%%%%%%%%%%%%%%%%%%%%%%
\item[(3,0)=10: rank-three symmetric]
%%%%%%%%%%%%%%%%%%%%%%%%%%%%%%%%%%%%%
\begin{align}
\chi_{\rm 3\mathchar`-sym.}
\equiv
    \chi_{(3,0)}
    =
    1+
    \lambda_1^3 + \lambda_2^3 + \lambda_3^3
    +
    \lambda_1^2\lambda_2 + \lambda_1^2\lambda_3 + \lambda_2^2\lambda_1 + \lambda_2^2\lambda_3 + \lambda_3^2\lambda_1 + 
    \lambda_3^2\lambda_2.
\end{align}
\end{description}
%%%%%%%%%%%%%%%%%%%%%%%%%%%%%%%%%%%%%
%%%%%%%%%%%%%%%%%%%%%%%%%%%%%%%%%%%%%
\subsection{Multiply-wound Polyakov loops and characters}\label{sec:winding-loop-vs-character}
%%%%%%%%%%%%%%%%%%%%%%%%%%%%%%%%%%%%%
%%%%%%%%%%%%%%%%%%%%%%%%%%%%%%%%%%%%%
The multiply-wound loops are expressed as
\begin{align}
    u_n 
    \equiv
    \frac{1}{N}\mathrm{Tr}P^n
    = 
    \frac{1}{N}\sum_{j=1}^N \lambda_j^n\, . 
    %=
    %\sum_r u_r^{(n)} \chi_r(\{\lambda\}).
\end{align}
For the $N=3$, the following relations hold:
\begin{align}
    3 u_1 &= \chi_{(1,0)} = \chi_{\rm fund.}\, ,
    \\
    3 u_2 
    &= 
    \chi_{(2,0)} - \chi_{(1,1)} = \chi_{\rm 2\mathchar`-sym.} - (\chi_{\rm fund.})^\ast\, ,
    \\
    3 u_3
    &=
    \chi_{(3,0)} - \chi_{(2,1)} + \chi_{(0,0)}
    =\chi_{\rm 3\mathchar`-sym.} - \chi_{\rm adj.}+ \chi_{\rm trivial}\, ,
    \\
    3 u_4
    &=
    \chi_{(4,0)} - \chi_{(3,1)} + \chi_{(1,0)}\, ,
    \\
    3 u_5
    &=
    \chi_{(5,0)} - \chi_{(4,1)} + \chi_{(2,0)}\, ,
    \\
    3 u_6
    &=
    \chi_{(6,0)} - \chi_{(5,1)} + \chi_{(3,0)}\, ,
    \\
  & \qquad\qquad\quad \vdots
   \nonumber
\end{align}
Note that only $u_3$ contains $\chi_\mathrm{trivial}$. 

Nonzero expectation values of $\chi_r$ for nontrivial representations $r$ characterize the deviation from the SU(3) Haar-random distribution. We emphasize that the Haar-randomness provides stronger restriction than that by the $\mathbb{Z}_3$ center symmetry: 
center symmetry allows nonzero values of $u_{3n}$.
The Haar-randomness, however, leads to $\chi_\mathrm{trivial} = 1$ and $\chi_r = 0$ for any other representation $r$, which allows only $u_3$ to be nonzero.
%%%%%%%%%%%%%
%\bibliographystyle{unsrt}
\bibliographystyle{utphys}
\bibliography{PD-letter}

\providecommand{\href}[2]{#2}\begingroup\raggedright\begin{thebibliography}{10}

\bibitem{Polyakov:1978vu}
A.~M. Polyakov, ``{Thermal Properties of Gauge Fields and Quark Liberation},''
  \href{http://dx.doi.org/10.1016/0370-2693(78)90737-2}{{\em Phys. Lett. B}
  {\bfseries 72} (1978) 477--480}.

\bibitem{Susskind:1979up}
L.~Susskind, ``{Lattice Models of Quark Confinement at High Temperature},''
  \href{http://dx.doi.org/10.1103/PhysRevD.20.2610}{{\em Phys. Rev. D}
  {\bfseries 20} (1979) 2610--2618}.

\bibitem{Sundborg:1999ue}
B.~Sundborg, ``{The Hagedorn transition, deconfinement and N=4 SYM theory},''
  \href{http://dx.doi.org/10.1016/S0550-3213(00)00044-4}{{\em Nucl. Phys. B}
  {\bfseries 573} (2000) 349--363},
  \href{http://arxiv.org/abs/hep-th/9908001}{{\ttfamily arXiv:hep-th/9908001}}.

\bibitem{Aharony:2003sx}
O.~Aharony, J.~Marsano, S.~Minwalla, K.~Papadodimas, and M.~Van~Raamsdonk,
  ``{The Hagedorn - deconfinement phase transition in weakly coupled large N
  gauge theories},'' \href{http://dx.doi.org/10.4310/ATMP.2004.v8.n4.a1}{{\em
  Adv. Theor. Math. Phys.} {\bfseries 8} (2004) 603--696},
  \href{http://arxiv.org/abs/hep-th/0310285}{{\ttfamily arXiv:hep-th/0310285}}.

\bibitem{Schnitzer:2004qt}
H.~J. Schnitzer, ``{Confinement/deconfinement transition of large N gauge
  theories with N(f) fundamentals: N(f)/N finite},''
  \href{http://dx.doi.org/10.1016/j.nuclphysb.2004.06.057}{{\em Nucl. Phys. B}
  {\bfseries 695} (2004) 267--282},
  \href{http://arxiv.org/abs/hep-th/0402219}{{\ttfamily arXiv:hep-th/0402219}}.

\bibitem{Hanada:2020uvt}
M.~Hanada, H.~Shimada, and N.~Wintergerst, ``{Color confinement and
  Bose-Einstein condensation},''
  \href{http://dx.doi.org/10.1007/JHEP08(2021)039}{{\em JHEP} {\bfseries 08}
  (2021) 039}, \href{http://arxiv.org/abs/2001.10459}{{\ttfamily
  arXiv:2001.10459 [hep-th]}}.

\bibitem{Hanada:2023rlk}
M.~Hanada and H.~Watanabe, ``{On thermal transition in QCD},''
  \href{http://arxiv.org/abs/2310.07533}{{\ttfamily arXiv:2310.07533
  [hep-th]}}.

\bibitem{Umeda:2012er}
{\bfseries WHOT-QCD} Collaboration, T.~Umeda, S.~Aoki, S.~Ejiri, T.~Hatsuda,
  K.~Kanaya, Y.~Maezawa, and H.~Ohno, ``{Equation of state in 2+1 flavor QCD
  with improved Wilson quarks by the fixed scale approach},''
  \href{http://dx.doi.org/10.1103/PhysRevD.85.094508}{{\em Phys. Rev. D}
  {\bfseries 85} (2012) 094508},
  \href{http://arxiv.org/abs/1202.4719}{{\ttfamily arXiv:1202.4719 [hep-lat]}}.

\bibitem{Aoki:2006br}
Y.~Aoki, Z.~Fodor, S.~D. Katz, and K.~K. Szabo, ``{The QCD transition
  temperature: Results with physical masses in the continuum limit},''
  \href{http://dx.doi.org/10.1016/j.physletb.2006.10.021}{{\em Phys. Lett. B}
  {\bfseries 643} (2006) 46--54},
  \href{http://arxiv.org/abs/hep-lat/0609068}{{\ttfamily
  arXiv:hep-lat/0609068}}.

\bibitem{Aoki:2006we}
Y.~Aoki, G.~Endrodi, Z.~Fodor, S.~D. Katz, and K.~K. Szabo, ``{The Order of the
  quantum chromodynamics transition predicted by the standard model of particle
  physics},'' \href{http://dx.doi.org/10.1038/nature05120}{{\em Nature}
  {\bfseries 443} (2006) 675--678},
  \href{http://arxiv.org/abs/hep-lat/0611014}{{\ttfamily
  arXiv:hep-lat/0611014}}.

\bibitem{Hanada:2019kue}
M.~Hanada and B.~Robinson, ``{Partial-Symmetry-Breaking Phase Transitions},''
  \href{http://dx.doi.org/10.1103/PhysRevD.102.096013}{{\em Phys. Rev. D}
  {\bfseries 102} no.~9, (2020) 096013},
  \href{http://arxiv.org/abs/1911.06223}{{\ttfamily arXiv:1911.06223
  [hep-th]}}.

\bibitem{Taniguchi:2016tjc}
Y.~Taniguchi, K.~Kanaya, H.~Suzuki, and T.~Umeda, ``{Topological susceptibility
  in finite temperature ( 2+1 )-flavor QCD using gradient flow},''
  \href{http://dx.doi.org/10.1103/PhysRevD.95.054502}{{\em Phys. Rev. D}
  {\bfseries 95} no.~5, (2017) 054502},
  \href{http://arxiv.org/abs/1611.02411}{{\ttfamily arXiv:1611.02411
  [hep-lat]}}.

\bibitem{Luscher:2010iy}
M.~L\"uscher, ``{Properties and uses of the Wilson flow in lattice QCD},''
  \href{http://dx.doi.org/10.1007/JHEP08(2010)071}{{\em JHEP} {\bfseries 08}
  (2010) 071}, \href{http://arxiv.org/abs/1006.4518}{{\ttfamily arXiv:1006.4518
  [hep-lat]}}. [Erratum: JHEP 03, 092 (2014)].

\bibitem{Hanada:2016pwv}
M.~Hanada and J.~Maltz, ``{A proposal of the gauge theory description of the
  small Schwarzschild black hole in AdS$_5\times$S$^5$},''
  \href{http://dx.doi.org/10.1007/JHEP02(2017)012}{{\em JHEP} {\bfseries 02}
  (2017) 012}, \href{http://arxiv.org/abs/1608.03276}{{\ttfamily
  arXiv:1608.03276 [hep-th]}}.

\bibitem{Berenstein:2018lrm}
D.~Berenstein, ``{Submatrix deconfinement and small black holes in AdS},''
  \href{http://dx.doi.org/10.1007/JHEP09(2018)054}{{\em JHEP} {\bfseries 09}
  (2018) 054}, \href{http://arxiv.org/abs/1806.05729}{{\ttfamily
  arXiv:1806.05729 [hep-th]}}.

\bibitem{Hanada:2018zxn}
M.~Hanada, G.~Ishiki, and H.~Watanabe, ``{Partial Deconfinement},''
  \href{http://dx.doi.org/10.1007/JHEP03(2019)145}{{\em JHEP} {\bfseries 03}
  (2019) 145}, \href{http://arxiv.org/abs/1812.05494}{{\ttfamily
  arXiv:1812.05494 [hep-th]}}. [Erratum: JHEP 10, 029 (2019)].

\bibitem{Hanada:2019czd}
M.~Hanada, A.~Jevicki, C.~Peng, and N.~Wintergerst, ``{Anatomy of
  Deconfinement},'' \href{http://dx.doi.org/10.1007/JHEP12(2019)167}{{\em JHEP}
  {\bfseries 12} (2019) 167}, \href{http://arxiv.org/abs/1909.09118}{{\ttfamily
  arXiv:1909.09118 [hep-th]}}.

\bibitem{Gross:1980he}
D.~J. Gross and E.~Witten, ``{Possible Third Order Phase Transition in the
  Large N Lattice Gauge Theory},''
\href{http://dx.doi.org/10.1103/PhysRevD.21.446}{{\em Phys. Rev.} {\bfseries
  D21} (1980) 446--453}.
%%CITATION = PHRVA,D21,446;%%.

\bibitem{Wadia:2012fr}
S.~R. Wadia, ``{A Study of U(N) Lattice Gauge Theory in 2-dimensions},''
\href{http://arxiv.org/abs/1212.2906}{{\ttfamily arXiv:1212.2906 [hep-th]}}.
%%CITATION = ARXIV:1212.2906;%%.

\bibitem{Hanada:2021ksu}
M.~Hanada, J.~Holden, M.~Knaggs, and A.~O'Bannon, ``{Global symmetries and
  partial confinement},'' \href{http://dx.doi.org/10.1007/JHEP03(2022)118}{{\em
  JHEP} {\bfseries 03} (2022) 118},
  \href{http://arxiv.org/abs/2112.11398}{{\ttfamily arXiv:2112.11398
  [hep-th]}}.

\bibitem{Buividovich:2015oju}
P.~V. Buividovich, G.~V. Dunne, and S.~N. Valgushev, ``{Complex Path Integrals
  and Saddles in Two-Dimensional Gauge Theory},''
  \href{http://dx.doi.org/10.1103/PhysRevLett.116.132001}{{\em Phys. Rev.
  Lett.} {\bfseries 116} no.~13, (2016) 132001},
  \href{http://arxiv.org/abs/1512.09021}{{\ttfamily arXiv:1512.09021
  [hep-th]}}.

\bibitem{Kaczmarek:2002mc}
O.~Kaczmarek, F.~Karsch, P.~Petreczky, and F.~Zantow, ``{Heavy quark anti-quark
  free energy and the renormalized Polyakov loop},''
  \href{http://dx.doi.org/10.1016/S0370-2693(02)02415-2}{{\em Phys. Lett. B}
  {\bfseries 543} (2002) 41--47},
  \href{http://arxiv.org/abs/hep-lat/0207002}{{\ttfamily
  arXiv:hep-lat/0207002}}.

\bibitem{Dumitru:2003hp}
A.~Dumitru, Y.~Hatta, J.~Lenaghan, K.~Orginos, and R.~D. Pisarski,
  ``{Deconfining phase transition as a matrix model of renormalized Polyakov
  loops},'' \href{http://dx.doi.org/10.1103/PhysRevD.70.034511}{{\em Phys. Rev.
  D} {\bfseries 70} (2004) 034511},
  \href{http://arxiv.org/abs/hep-th/0311223}{{\ttfamily arXiv:hep-th/0311223}}.

\bibitem{Gupta:2006qm}
S.~Gupta, K.~Huebner, and O.~Kaczmarek, ``{Polyakov loop in different
  representations of SU(3) at finite temperature},''
  \href{http://dx.doi.org/10.1016/j.nuclphysa.2006.11.160}{{\em Nucl. Phys. A}
  {\bfseries 785} (2007) 278--281},
  \href{http://arxiv.org/abs/hep-lat/0608014}{{\ttfamily
  arXiv:hep-lat/0608014}}.

\bibitem{Gupta:2007ax}
S.~Gupta, K.~Huebner, and O.~Kaczmarek, ``{Renormalized Polyakov loops in many
  representations},'' \href{http://dx.doi.org/10.1103/PhysRevD.77.034503}{{\em
  Phys. Rev. D} {\bfseries 77} (2008) 034503},
  \href{http://arxiv.org/abs/0711.2251}{{\ttfamily arXiv:0711.2251 [hep-lat]}}.

\bibitem{Mykkanen:2012ri}
A.~Mykkanen, M.~Panero, and K.~Rummukainen, ``{Casimir scaling and
  renormalization of Polyakov loops in large-N gauge theories},''
  \href{http://dx.doi.org/10.1007/JHEP05(2012)069}{{\em JHEP} {\bfseries 05}
  (2012) 069}, \href{http://arxiv.org/abs/1202.2762}{{\ttfamily arXiv:1202.2762
  [hep-lat]}}.

\bibitem{Petreczky:2015yta}
P.~Petreczky and H.~P. Schadler, ``{Renormalization of the Polyakov loop with
  gradient flow},'' \href{http://dx.doi.org/10.1103/PhysRevD.92.094517}{{\em
  Phys. Rev. D} {\bfseries 92} no.~9, (2015) 094517},
  \href{http://arxiv.org/abs/1509.07874}{{\ttfamily arXiv:1509.07874
  [hep-lat]}}.

\bibitem{Bergner:2023rpw}
G.~Bergner, V.~Gautam, and M.~Hanada, ``{Color confinement and random matrices.
  A random walk down group manifold toward Casimir scaling},''
  \href{http://dx.doi.org/10.1007/JHEP03(2024)013}{{\em JHEP} {\bfseries 03}
  (2024) 013}, \href{http://arxiv.org/abs/2311.14093}{{\ttfamily
  arXiv:2311.14093 [hep-th]}}.

\bibitem{Asakawa:1995zu}
M.~Asakawa and T.~Hatsuda, ``{What thermodynamics tells about QCD plasma near
  phase transition},'' \href{http://dx.doi.org/10.1103/PhysRevD.55.4488}{{\em
  Phys. Rev. D} {\bfseries 55} (1997) 4488--4491},
  \href{http://arxiv.org/abs/hep-ph/9508360}{{\ttfamily arXiv:hep-ph/9508360}}.

\bibitem{Glozman:2022zpy}
L.~Y. Glozman, ``{Chiral spin symmetry and hot/dense QCD},''
  \href{http://dx.doi.org/10.1016/j.ppnp.2023.104049}{{\em Prog. Part. Nucl.
  Phys.} {\bfseries 131} (2023) 104049},
  \href{http://arxiv.org/abs/2209.10235}{{\ttfamily arXiv:2209.10235
  [hep-lat]}}.

\bibitem{Cohen:2023hbq}
T.~D. Cohen and L.~Y. Glozman, ``{Large $N_c$ QCD phase diagram at $\mu_B =
  0$},'' \href{http://arxiv.org/abs/2311.07333}{{\ttfamily arXiv:2311.07333
  [hep-ph]}}.

\bibitem{Gross:1993hu}
D.~J. Gross and W.~Taylor, ``{Two-dimensional QCD is a string theory},''
  \href{http://dx.doi.org/10.1016/0550-3213(93)90403-C}{{\em Nucl. Phys. B}
  {\bfseries 400} (1993) 181--208},
  \href{http://arxiv.org/abs/hep-th/9301068}{{\ttfamily arXiv:hep-th/9301068}}.

\bibitem{Gross:1993yt}
D.~J. Gross and W.~Taylor, ``{Twists and Wilson loops in the string theory of
  two-dimensional QCD},''
  \href{http://dx.doi.org/10.1016/0550-3213(93)90042-N}{{\em Nucl. Phys. B}
  {\bfseries 403} (1993) 395--452},
  \href{http://arxiv.org/abs/hep-th/9303046}{{\ttfamily arXiv:hep-th/9303046}}.

\bibitem{Berenstein:2023srv}
D.~Berenstein and K.~Yan, ``{The endpoint of partial deconfinement},''
  \href{http://dx.doi.org/10.1007/JHEP12(2023)030}{{\em JHEP} {\bfseries 12}
  (2023) 030}, \href{http://arxiv.org/abs/2307.06122}{{\ttfamily
  arXiv:2307.06122 [hep-th]}}.

\bibitem{einstein1924quantentheorie}
A.~Einstein, ``Quantentheorie des einatomigen idealen gases,'' {\em S-B Preuss.
  Akad. Berlin} (1924) .

\bibitem{Nishigaki:2024phx}
S.~Nishigaki, ``{Eigenphase distributions of unimodular circular ensembles},''
  \href{http://arxiv.org/abs/2401.09045}{{\ttfamily arXiv:2401.09045
  [math-ph]}}.

\end{thebibliography}\endgroup

\end{document}